%% file: CNF-2024-v8.tex
\documentclass{elsarticle}
\usepackage{lineno,hyperref}

\modulolinenumbers[5]

\usepackage{siunitx}
\usepackage{placeins}
\usepackage{multirow}
\usepackage{lscape}
\usepackage{amsmath}
\usepackage{amssymb}
\usepackage{caption}
\usepackage[list=true,labelformat=simple]{subcaption}
\usepackage{subcaption}
\usepackage{multirow}
\usepackage{array}

\usepackage{graphicx}

\PassOptionsToPackage{breaklinks=true}{hyperref}
\usepackage{hyperref}
\usepackage[breaklinks=true]{hyperref}
\usepackage{url}  

\makeatletter
\def\ps@pprintTitle{%
 \let\@oddhead\@empty
 \let\@evenhead\@empty
 \let\@oddfoot\@empty
 \let\@evenfoot\@empty
}
\makeatother

\begin{document}

\begin{frontmatter}

\title{Role of the argon and helium bath gases on the structure of $\text{H}_2$/$\text{O}_2$ detonations}


\author{Farzane Zangene\corref{mycorrespondingauthor}}
\ead{fzang055@uottawa.ca}
\author{Matei I. Radulescu}
\address{Department of Mechanical Engineering, University of Ottawa, Ottawa, ON K1N6N5, Canada}




\begin{abstract}

This study investigates the role of two inert mono-atomic diluents, argon and helium, on the detonation structure in order to assess the importance of vibrational non-equilibrium and wall losses.  When relaxation effects and wall losses are neglected, the detonation waves in mixtures diluted with either of these gases have the same kinetics, Mach number, and specific heat ratio and hence are expected to lead to the same cellular dynamics. Differences in transport properties and species relaxation rates thus permit to establish the importance of these effects.  The experiments were conducted in $2\text{H}_2+\text{O}_2+7\text{Ar}$ and $2\text{H}_2+\text{O}_2+7\text{He}$ mixtures in a narrow channel, where boundary layer losses can be controlled by the proximity of the detonations to their propagation limits. The initial pressure was adjusted in such a way that the induction zone length (therefore cell sizes) calculated from the ideal ZND model remained constant.  The experiments revealed differences in velocity deficits and cell sizes despite maintaining a constant induction zone length across the mixtures. These differences were minimal in sensitive mixtures but became more pronounced as velocity deficits increased and cell sizes approached the channel dimensions. Near the detonation limits, the disparity in cell sizes between the two mixtures nearly doubled.  These observations were reconciled by accounting for the wall losses.  We incorporated the boundary layer flow divergence in a perturbation analysis based on the square wave detonation assumption.  This permitted to establish the controlling loss parameter as the product of the induction to channel size and the inverse of the square root of the Reynolds number.  The very good collapse of the scaled results with the two bath gases with the loss parameter, and further comparison with 2D numerical simulations with account for flow divergence to the third dimension, confirmed the viscous loss mechanism to be dominating. Calculations suggest that the slower relaxation of $\text{H}_2$ becomes comparable with the ignition delay anticipated from the ZND model and is slower by 70\% in the argon diluted system.  Differences possibly highlighting the role of non-equilibrium were not observed.  This suggests the vibrational non-equilibrium effect may be less apparent in cellular detonations due to the lengthening of the ignition delays owing to the non-steady detonation structure.  The study establishes that the large differences between the enlarged cells observed in our experiments and numerical predictions of lossless systems can be entirely attributed to wall losses.  

\end{abstract}

\begin{keyword}
\texttt{Detonation cellular structure, Vibrational relaxation, Boundary layer losses, Inert diluent}
\end{keyword}

\end{frontmatter}


\section{Introduction}
\label{sec:intro}
The cellular structure of gaseous detonation waves has been known for over 60 years \cite{white1961turbulent, denisov1959pulsating}. Qualitatively, the physico-chemical processes occurring in the detonation structure are well understood, particularly for systems displaying regular cells \cite{Fickett:1979}.  Nevertheless, recent attempts to quantitatively model the detonation structure have led to large discrepancies between the numerical predictions and experimentally determined cell sizes \cite{taylor2013numerical}. These differences suggest the importance of other effects neglected in the modelling. Taylor et al.\ attribute these effects to possible influence of vibrational non-equilibrium \cite{taylor2013numerical}.  They suggest that the excess translational energy in the induction zone may promote the rate processes in the induction zone chemical kinetics and account for the smaller cells. On the other hand, another possible explanation was provided by Xiao et al.\ \cite{Xiao:2021}.  They suggest that wall losses may have a strong influence on the detonation cells:  with larger velocity deficits, the kinetics are slowed down and may yield much larger cells than detonations propagating without wall losses, as calculated by Taylor et al \cite{taylor2013numerical}.  Indeed, experimental measurements of cell sizes reported in the literature are never corrected for the velocity deficit effects, hence do not permit direct comparison with lossless calculations.  To this date, limited attempts have been made to answer this question.  The preliminary experiments of this work were presented at the International Conference on Hydrogen Safety in 2021 \cite{zangene2021role} and the modelling in 2022 at the International Colloquium on the Dynamics of Explosions and Reactive Systems \cite{ZangeneIcder:2022}.  Subsequently, Shi et al.\ reported experiments and analysis of the possible speed-up in the induction zone owing to the vibrational non-equilibrium, but found the effect to be negligible \cite{shi2024probing}. Other changes in cellular dynamics have also been suggested by analysing the changes in the ZND rate processes in a simplified toy model \cite{uy2018chemical}.  On the other hand, Smith et al.\ conducted numerical simulations of $2\text{H}_2+\text{O}_2+7\text{Ar}$ detonations in thin channels, accounting for the losses in an improved model over that of Xiao et al. \cite{Xiao:2021} with a higher fidelity chemical model, but neglecting vibrational non-equilibrium, and found good agreement with experiment \cite{smith2024nature}.  
\begin{figure}[t]
\centering
\begin{subfigure}{0.48\textwidth}
    \includegraphics[scale=0.35]{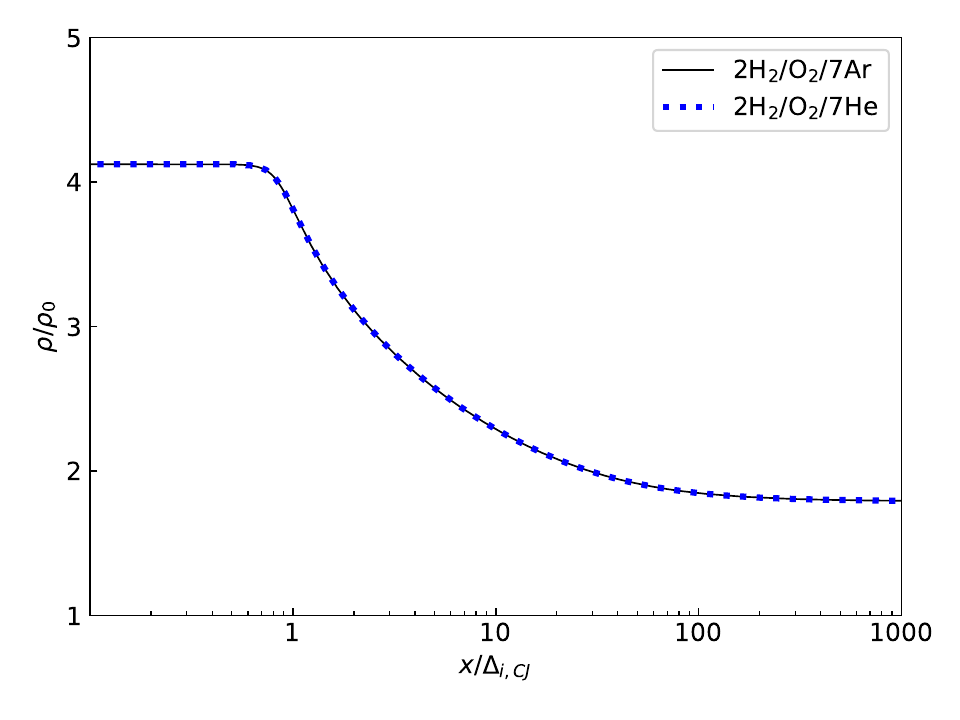}
\label{fig:ZND-H-Tx}
\end{subfigure} 
\begin{subfigure}{0.48\textwidth}
    \includegraphics[scale=0.35]{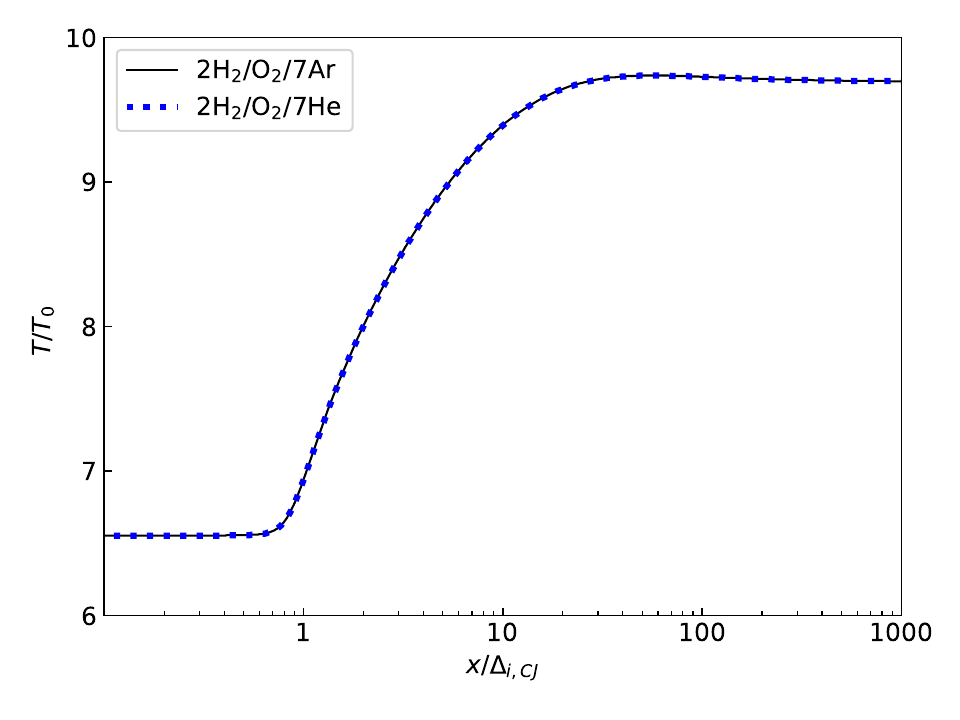}
\label{fig:ZND-H-Tx}
\end{subfigure} 
\caption{ZND profiles of (a) density and (b) temperature obtained by using the San Diego detailed chemical mechanism \cite{sandiego} at $p_0$= 10 \si{\kilo\pascal} and $T_0$= 293 \si{\kelvin}. $\Delta_{i, CJ}$ is the induction zone length calculated from the ideal ZND model, $\rho_0$ is the initial density of the mixture.}
\label{fig:idealZND}
\end{figure}
In the current communication, we wish to establish the relative role of these two effects experimentally.  The strategy we use is the dilution of the $2\text{H}_2+\text{O}_2$ system with either argon gas or helium gas.   The diluents being chemically inert and mono-atomic, when relaxation effects and wall losses are neglected, the detonation waves in mixtures diluted with either of these gases have the same kinetics, Mach number, and specific heat ratio and hence are expected to lead to the same cellular dynamics. Differences in transport properties and species relaxation rates hence permit to establish the importance of these effects in a non-ambiguous manner. The experiments are conducted in a narrow channel, such that the magnitude of wall losses can be controlled by the proximity of the operating conditions to the minimum channel thickness permitting propagation. 

For the same amount of dilution, initial pressure, and temperature, the diluent does not alter the ignition delay time since the shock Mach number and temperature ($T_{VN}$:Von-Neumann state) remain constant. Additionally, the effective activation energy ($E_a/RT_{VN}$) and the time scale ratio of induction to reaction ($t_i/t_r$) are unaffected. Consequently, no significant differences in the instability parameter, $\chi = \left(t_i/t_r\right) \left(E_a/RT_{VN}\right)$, are expected \cite{radulescu2003propagation}. Thermodynamically, the choice of diluent does not influence the specific heat ratio ($\gamma$) or heat release, resulting in the same ZND structure, as shown in Fig.\ \ref{fig:idealZND}. However, the relaxation rate of the reactants H$_2$ and O$_2$ will be affected by the type of diluent due to their significant difference in molecular weight. This will permit us to isolate their effect on the detonation structure. 

Previous studies on the nature of the inert diluent are limited. Experiments in propane mixtures using both of these diluents as bath gases \cite{haloua2000characteristics} in thin tubes demonstrated that galloping detonations were established near the propagation limits more readily in argon diluted mixtures than in helium diluted mixtures.  This observation has not been resolved, but point to the importance of either the differences in transport coefficients in the helium mixture (due to increased thermal conductivity and sound speed), which may enhance dissipation and suppress the fine-scale instabilities necessary for the galloping phenomenon, or the vibrational non-equilibrium effects.  The latter would act in the same direction, with non-equilibria being more important in the argon based system.  Kumar et al.\ measured the cell widths experimentally in hydrogen-oxygen-helium and hydrogen-oxygen-argon mixtures at the same initial pressures and temperatures and showed that the cells in argon-diluted mixtures are smaller than those in helium-diluted mixtures  \cite{kumar1990detonation, kumar1995detonation}. This difference was attributed to the variations in molecular weights (or post-shock velocities) between the helium and argon mixtures. Indeed, while ZND structure is the same in terms of time, length scales with the acoustic speed, will changes due to the difference in molecular weights.  For a meaningful comparison, our experiments fix the induction zone length by adjusting the initial pressure. 

The paper is organized as follows: Section \ref{sec:Exp} outlines the experimental procedure and presents the results. Section \ref{sec:Vib} estimates the vibrational non-equilibrium time in the two mixtures with different bath gases, comparing it to the ignition delay time of each mixture. Section \ref{sec:Num} discusses the modifications made to the existing boundary layer model, presents the numerical simulation results for detonation propagation, and compares these outcomes with experimental data. In the same section, the results of the zero-dimensional ZND model with and without the losses are provided, followed by a discussion. Finally, section \ref{sec:Con} provides a general discussion, followed by the study's conclusions.  The appendix provides a closed model for detonations with boundary layer losses assuming the structure as a square wave.  Through a perturbation analysis in the limit of high activation energy, we derive the dependence of the detonation speed on a newly derived  loss parameter involving the ratio of induction to channel size and the Reynolds number.

\begin{figure}[t]
\centering
\includegraphics[scale=0.82]{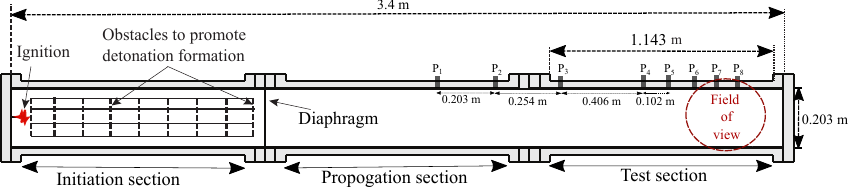}
\caption{Schematic of the shock tube used in the experiment.}
\label{fig:ST}
\end{figure}

\section{Experiments}
\label{sec:Exp}
\subsection{Experimental details} 
Experiments were conducted in a rectangular shock tube with dimensions of 3.4 \si{\meter} in length, 0.203 \si{\meter} in height, and 0.019 \si{\meter} in width. The channel, consisting of initiation, propagation, and test sections, was made of aluminum. The schematic illustrating the experimental set-up is presented in Fig.\ \ref{fig:ST}, with additional details available in a previous work \cite{Bhattacharjee}. The test section of the shock tube features glass walls for visualization. A Z-type schlieren set-up, comprising a slit, a vertical knife-edge, an incandescent 360 Watt lamp, and two concave mirrors, was employed to visualize the propagation of the detonation at the end of the test section by capturing the refraction of light rays. The recording was done at 77481 frames per second, with an exposure time of 0.47 \si{\micro\second} and a resolution of 384 \(\times\) 288 $\text{px}^2$ using a Phantom v1210 camera. Additionally, to visualize the detonation evolution process along the entire test section, 1 \si{\meter} long, a large-scale shadow-graph system was used. This system featured a 2 m \(\times\) 2 m retro-reflective screen and a 1000 W Xenon arc lamp. The high-speed camera had a resolution of 1024 \(\times\) 272 $\text{px}^2$, a frame rate of 44000, and an exposure time of 0.47 \si{\micro\second}.
\begin{table*}[t]
\centering
\caption{Experiments test gases with their ideal ZND prediction.}
\begin{tabular}{|c|c|c|c|}
\hline
Test gases          & $p_0$ [\si{\kilo\pascal}] & $\Delta_{i, CJ}$ [\si{\milli\meter}] & $D_{CJ}$ [\si{\meter\per\second}] \\ 
\hline
\multirow{2}{*}{$\text{2H}_\text{2}/\text{O}_\text{2}/7\text{Ar}$} & 4.1   &  2.8 &  1602 \\ \cline{2-4}
                                 & 7.2   &  1.6 &  1618 \\ 
\hline
\multirow{3}{*}{$\text{2H}_\text{2}/\text{O}_\text{2}/7\text{He}$} & 6.6   &  4   &  3588 \\ \cline{2-4}
                                 & 9.3   &  2.8 &  3609 \\ \cline{2-4}
                                 & 15    &  1.6 &  3693 \\ 
\hline
\end{tabular}
\label{tab:tests}
\end{table*}

\begin{figure}[t]
\centering
\begin{subfigure}{0.48\textwidth}
    \includegraphics[scale=0.35]{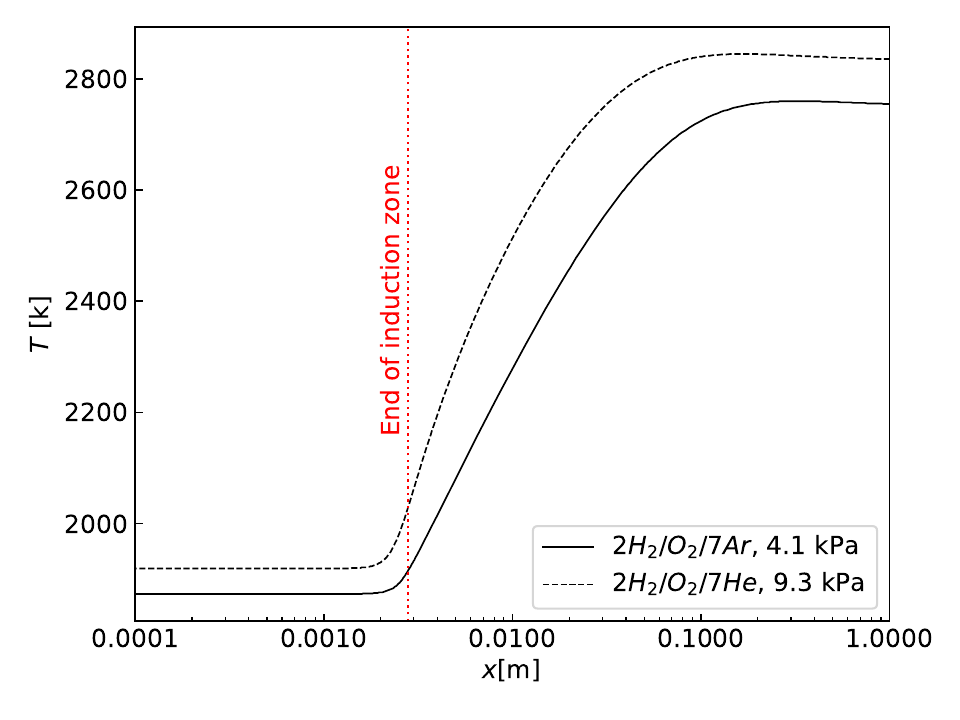}
    \label{fig:ZND-L-Tx}
\end{subfigure}
\begin{subfigure}{0.48\textwidth}
    \includegraphics[scale=0.35]{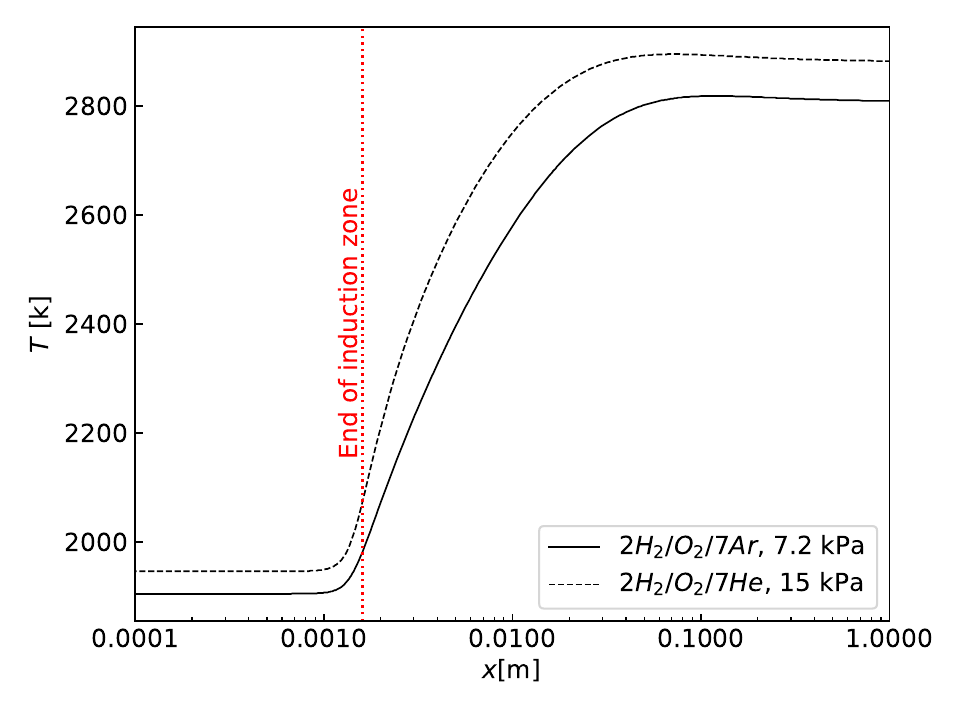}
\label{fig:ZND-H-Tx}
\end{subfigure} 
\caption{ZND profiles of temperature obtained as a function of distance using the San Diego detailed chemical mechanism \cite{sandiego} at $T_0$ = 293 K.}
\label{fig:ZND}
\end{figure}

The two mixtures studied were a stoichiometric mixture of hydrogen-oxygen diluted with argon ($\text{2H}_\text{2}/\text{O}_\text{2}/7\text{Ar}$) and a stoichiometric mixture of hydrogen-oxygen diluted with helium ($\text{2}\text{H}_\text{2}/\text{O}_\text{2}/\text{7}\text{He}$). 
The initial pressure, $p_0$, of the test gases was adjusted such that the induction zone length, $\Delta_{i,CJ}$, calculated from the ideal Zel'dovich–Von-Neumann–Doering (ZND) model \cite{zel1940theory, von1943theory, doring1943detonation}, is kept constant between the two different mixtures. Considering the classical correlation between cell width, $\lambda$, and induction zone length \cite{westbrook1982chemical, Lee:2008}, ideally, we expect to observe the same cellular structure between the two mixtures.  Table \ref{tab:tests} shows the experimental test gases, their corresponding induction zone length and Chapman-Jouguet (CJ) detonation speed, $D_{CJ}$, at each initial pressure. Furthermore, Fig.\ \ref{fig:ZND} shows the corresponding ZND structures of the two mixtures at different initial pressures. The ideal ZND model prediction is calculated using a Python code working under the framework of SDToolbox \cite{sandiego} and Cantera \cite{cantera}. The point in the post-shock region in which thermicity is maximum is considered as the end of the induction zone.  In the lower pressure cases (4.1 \si{\kilo\pascal} for argon diluted and 9.3 \si{\kilo\pascal} for helium diluted) for each mixture, the measured induction zone length is 2.8-mm, and it is 1.6-mm in the higher pressure cases (7.2 \si{\kilo\pascal} for argon diluted and 15 \si{\kilo\pascal} for helium diluted mixture).
Additional experiments were conducted at a lower pressure of 6.6 \si{\kilo\pascal} in the hydrogen-oxygen-helium mixture to reduce the mixture's sensitivity and improve visualization of the detonation propagation.
\begin{figure}[t!]
\centering
\begin{subfigure}{0.9\textwidth}
    \includegraphics[width=\textwidth]{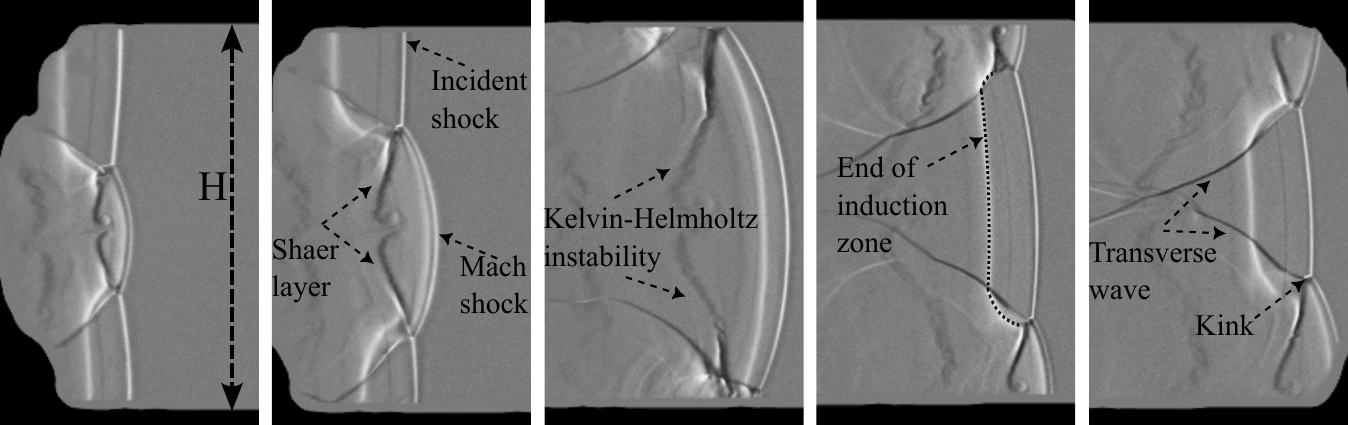}
    \caption{Schlieren imaging of detonation in the mixture of $\text{2H}_\text{2}/\text{O}_\text{2}/7\text{Ar}$ at $p_0$= 4.1 \si{\kilo\pascal}.}
    \label{fig:Ar4Sch}
\end{subfigure}
\begin{subfigure}{0.9\textwidth}
    \includegraphics[width=\textwidth]{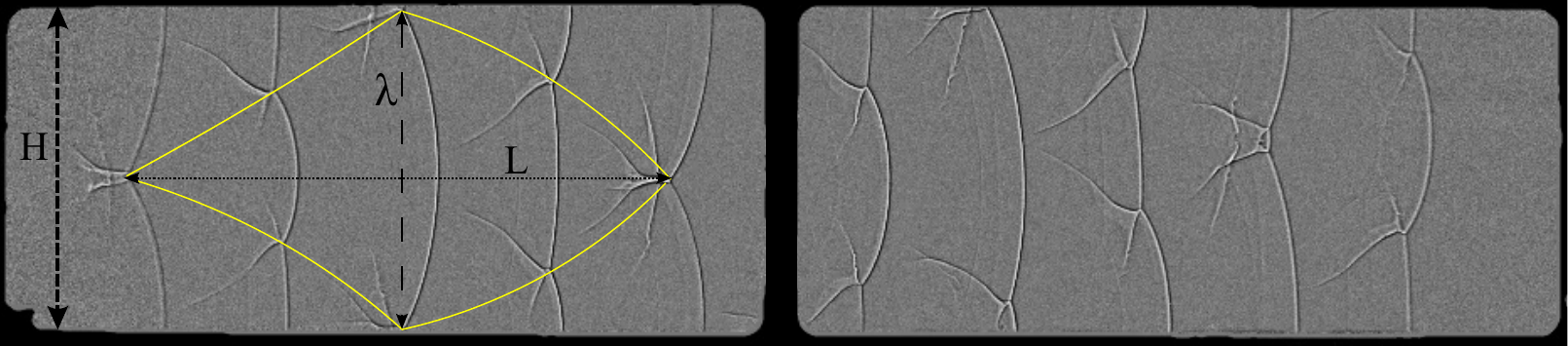}
    \caption{Shadowgraph imaging of detonation in the mixture of $\text{2H}_\text{2}/\text{O}_\text{2}/7\text{Ar}$ at $p_0$= 4.1 \si{\kilo\pascal}.} 
\label{fig:Ar4Shad}
\end{subfigure} 
\begin{subfigure}{0.9\textwidth}
    \includegraphics[width=\textwidth]{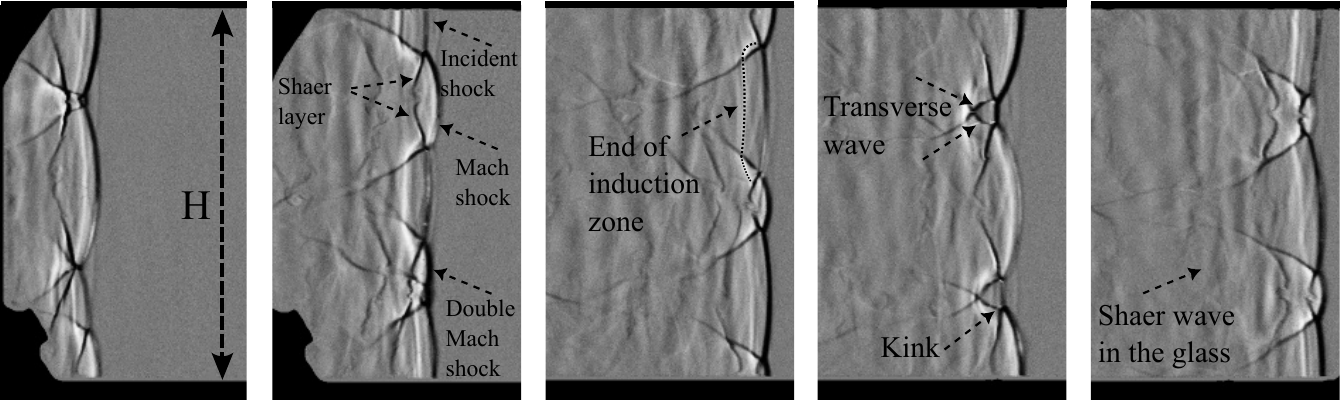}
    \caption{Schlieren imaging of detonation in the mixture of $\text{2H}_\text{2}/\text{O}_\text{2}/7\text{He}$ at $p_0$= 9.3 \si{\kilo\pascal}.}
    \label{fig:He9Sch}
\end{subfigure}
\begin{subfigure}{0.9\textwidth}
    \includegraphics[width=\textwidth]{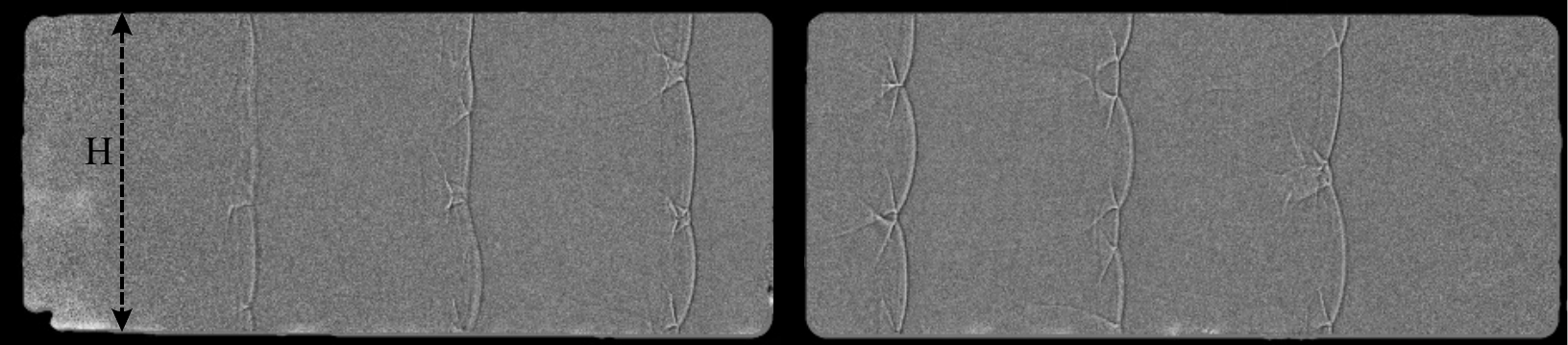}
    \caption{Shadowgraph imaging of detonation in the mixture of $\text{2H}_\text{2}/\text{O}_\text{2}/7\text{He}$ at $p_0$= 9.3 \si{\kilo\pascal}.}
\label{fig:}
\end{subfigure} 
\caption{Detonation propagating from left to right in mixtures with a fix induction zone length of $\Delta_i = 2.8$ \si{\milli\meter} at $T_0 = 293$ K. The visualization length is 300 \si{\milli\meter} using the Schlieren technique and 1000 \si{\milli\meter} using the shadowgraph technique. The channel height $H$ is 203 \si{\milli\meter}, $L$ is the cell length and $\lambda$ is the cell width.}
\label{fig:expArHelow}
\end{figure}
\begin{figure}[t!]
\centering
\begin{subfigure}{0.9\textwidth}
    \includegraphics[width=\textwidth]{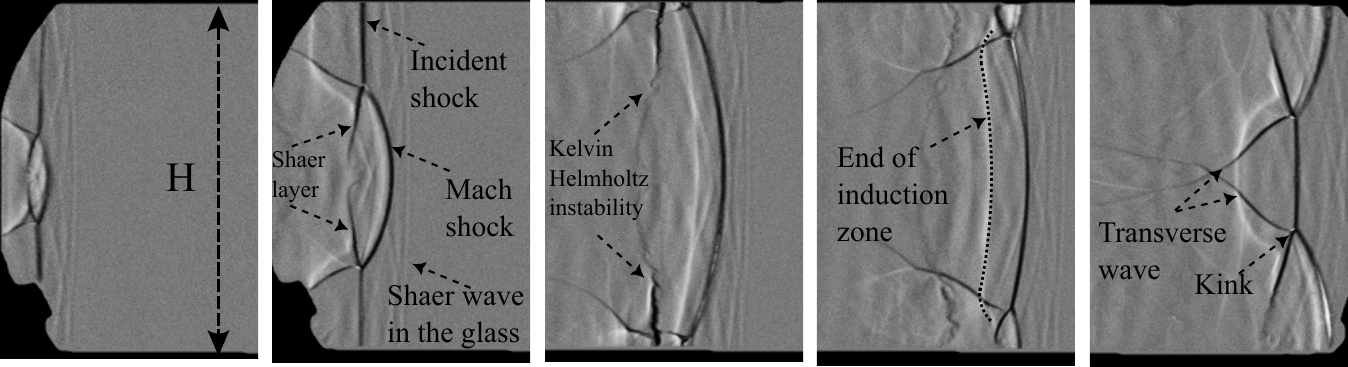}
    \caption{Schlieren imaging of detonation in the mixture of $\text{2H}_\text{2}/\text{O}_\text{2}/7\text{He}$ at $p_0$= 6.6 \si{\kilo\pascal}.} 
    \label{fig:He6Sch}
\end{subfigure}
\begin{subfigure}{0.9\textwidth}
    \includegraphics[width=\textwidth]{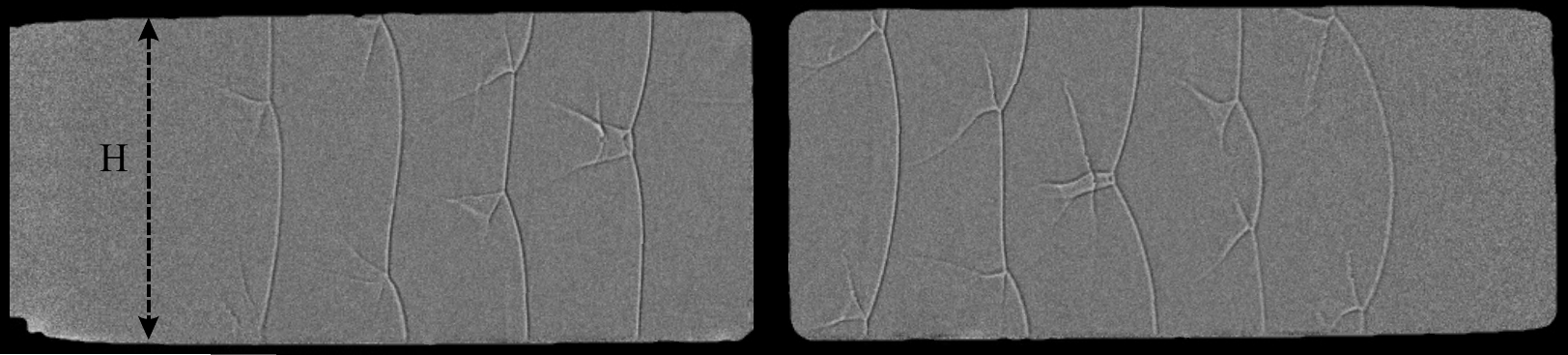}
    \caption{Shadowgraph imaging of detonation in the mixture of $\text{2H}_\text{2}/\text{O}_\text{2}/7\text{He}$ at $p_0$= 6.6 \si{\kilo\pascal}.}
\label{fig:}
\end{subfigure} 
\caption{Detonation propagating from left to right in mixture with an induction zone length of $\Delta_i = 4$ \si{\milli\meter} at $T_0 = 293$ K. The visualization length is 300 \si{\milli\meter} using the Schlieren technique and 1000 \si{\milli\meter} using the shadowgraph technique. The channel height $H$ is 203 \si{\milli\meter}.}
\label{fig:expHe6}
\end{figure}

\begin{figure}[t!]
\centering
\begin{subfigure}{0.9\textwidth}
    \includegraphics[width=\textwidth]{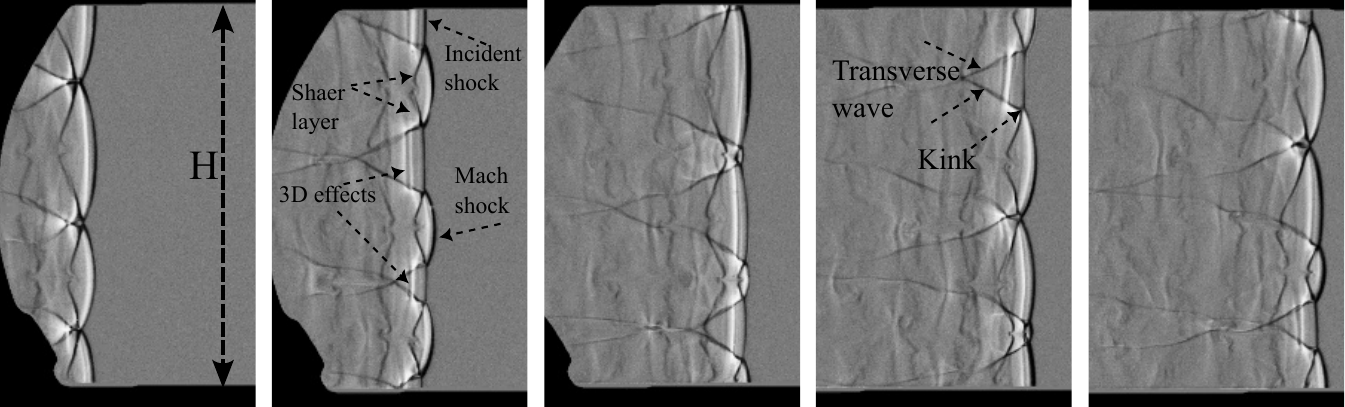}
    \caption{Schlieren imaging of detonation in the mixture of $\text{2H}_\text{2}/\text{O}_\text{2}/7\text{Ar}$ at $p_0$= 7.2 \si{\kilo\pascal}.}
    \label{fig:Ar7Sch}
\end{subfigure}
\begin{subfigure}{0.9\textwidth}
    \includegraphics[width=\textwidth]{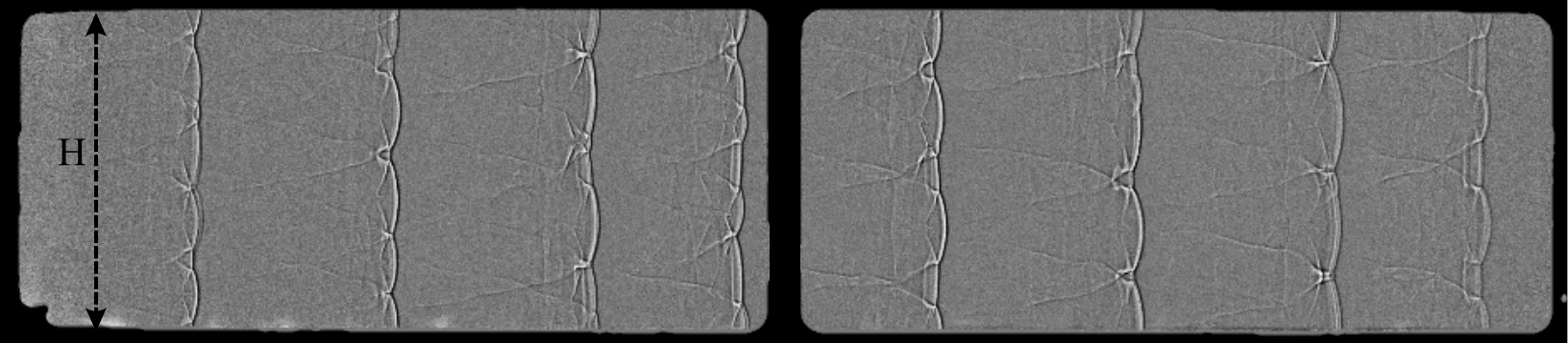}
    \caption{Shadowgraph imaging of detonation in the mixture of $\text{2H}_\text{2}/\text{O}_\text{2}/7\text{Ar}$ at $p_0$= 7.2 \si{\kilo\pascal}.} 
\label{fig:Ar7shad}
\end{subfigure} 
\begin{subfigure}{0.9\textwidth}
    \includegraphics[width=\textwidth]{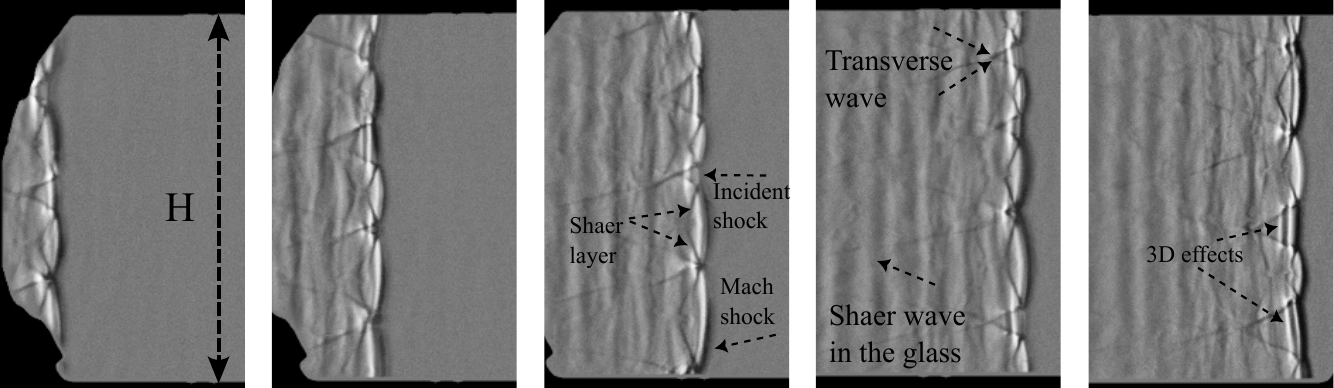}
    \caption{Schlieren imaging of detonation in the mixture of $\text{2H}_\text{2}/\text{O}_\text{2}/7\text{He}$ at $p_0$= 15 \si{\kilo\pascal}.}
    \label{fig:}
\end{subfigure}
\begin{subfigure}{0.9\textwidth}
    \includegraphics[width=\textwidth]{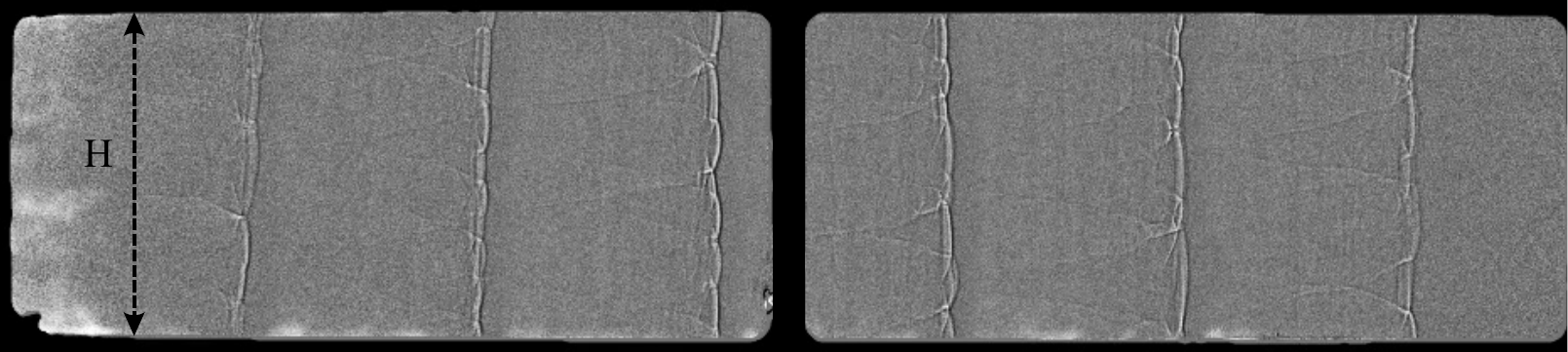}
    \caption{Shadowgraph imaging of detonation in the mixture of $\text{2H}_\text{2}/\text{O}_\text{2}/7\text{He}$ at $p_0$= 15 \si{\kilo\pascal}.} 
\label{fig:}
\end{subfigure} 
\caption{Detonation propagating from left to right in mixtures with a fix induction zone length of $\Delta_i = 1.6$ \si{\milli\meter} at $T_0 = 293$ K. The visualization length is 300 \si{\milli\meter} using the Schlieren technique and 1000 \si{\milli\meter} using the shadowgraph technique. The channel height $H$ is 203 \si{\milli\meter}.}
\label{fig:expArHeHigh}
\end{figure}

Each mixture was prepared in a separate mixing tank using the method of partial pressures and was left to mix for more than 24 hours. Before filling with driver and test gases, the shock tube was evacuated to a pressure below 80 Pa. Initial ignition of the premixed combustible mixture was achieved using a custom-designed and built high voltage igniter (HVI) \cite{Bhattacharjee}, which can store up to 1000 \si{\joule} of energy with a deposition time of 2 \si{\micro\second}. In all experiments, a stoichiometric mixture of ethylene-oxygen (\(\text{C}_2\text{H}_4/\text{O}_2\)) was used as the driver gas to initiate detonation. To avoid an over-driven detonation wave in the test section due to the high-pressure driver gas, numerous pilot tests were conducted to determine the lowest pressure of the driver gas capable of initiating detonation waves. The ratio of 2.5 between driver gas and test gas was maintained consistently across all experiments as the reference for filling the shock tube. Experiments in the hydrogen-oxygen-argon mixture were repeated five times, and in the hydrogen-oxygen-helium mixture ten times, given the reduction in data obtained from a single experiment due to the higher propagation speed  in this system. 

Eight high-frequency piezoelectric PCB pressure sensors (models 113B24 and 113B27) were installed on the top wall of the shock tube to record pressure signals and determine propagation speeds using the time-of-arrival method. The sensors resonate at 500 \si{\kilo\hertz}, and the pressure signals during the experiments were recorded at a rate of 1.5 \si{\mega\hertz}. All pressure gauges used in the experiments have a diameter of 5.5 mm and a maximum error of 1.3\%, as determined from the calibration data.

\subsection{Experimental results} 

Figure \ref{fig:Ar4Sch} presents the results of schlieren and shadowgraph methods for both argon- and helium-diluted mixtures, with a fixed $\Delta_{i,CJ}$ of 2.8 mm. The schlieren images depict a sequence of frames capturing the detonation wave as it propagates from left to right across a 30 cm distance (matching the diameter of the mirror) at the end of the test section. The superimposed shadowgraph images illustrate the evolution of detonation fronts along the entire 1-meter test section. For the argon-diluted mixture shown in Fig.\ \ref{fig:Ar4Sch}, a typical regular cellular structure of the detonation is observed. Detonation exhibits significantly enlarged cellular structure, with the cell height constrained by the channel dimensions, indicating mode locking. At the low initial pressure of 4.1 \si{\kilo\pascal}, the detonation has relatively large unburned induction zones, visible behind the incident shock.  The end of the induction zone is readily established by the marked density gradient in the schlieren photographs associated with the density drop upon exothermicity.  The induction zone terminates within a close distance behind the transverse shocks.  Vortex structures characteristic of Kelvin–Helmholtz instability are also clearly visible along the shear layer emanating from the triple points.  The structure of the triple point is characteristic of the reactive transverse wave configuration previously observed in these regular systems - see Fickett and Davis for review \cite{Fickett:1979}.   

Figure \ref{fig:Ar4Shad} displays the shadowgraph result obtained over a much larger length scale.  The same dynamics are observed, but some details are not as apparent, such that the location of the end of the induction zone.  The results confirm that the single-cell structure is maintained throughout the entire propagation in the test section. 

Figure \ref{fig:He9Sch} shows the detonation structure observed in $\text{2H}_\text{2}/\text{O}_\text{2}/7\text{He}$ mixture at the same nominal induction length. Although the induction zone length is fixed, the cell sizes are approximately half as large compared to those in the argon-diluted mixtures.  The local dynamics appear similar on a lengths scale of a cell.  The induction zone behind the leading shock has the same configuration, terminating at the same scaled distance behind the lead shock and transverse waves.  

The results obtained with helium dilution are nevertheless characterized by a system of vertical striations, absent in the argon experiments, as seen in Fig.\ \ref{fig:He9Sch}.  In some cases, this pattern outruns the detonation front (Fig.\ \ref{fig:He6Sch}).  We believe these are shear waves propagating in glass walls confining the detonation.  In the helium mixture, the detonation speed (nominally $D_{CJ}\sim$ 3600 \si{\meter/\second}) is very close to the shear wave in glass (3764 \si{\meter/\second} \cite{lide2004crc}).

To better visualize the structural details in the helium-diluted mixture, the initial pressure was reduced to 6.6 \si{\kilo\pascal}, resulting in one large detonation cell, as shown in Fig.\ \ref{fig:expHe6}. The structure's details, including the presence of Kelvin-Helmholtz instabilities along the shear layer, are similar to those observed in the single large cell of the argon-diluted mixture, although the instabilities appear less prominent in the helium-diluted mixture. 
\begin{figure}[t!]
    \centering
    \begin{subfigure}[b]{0.75\textwidth}
        \centering
        \includegraphics[width=\textwidth]{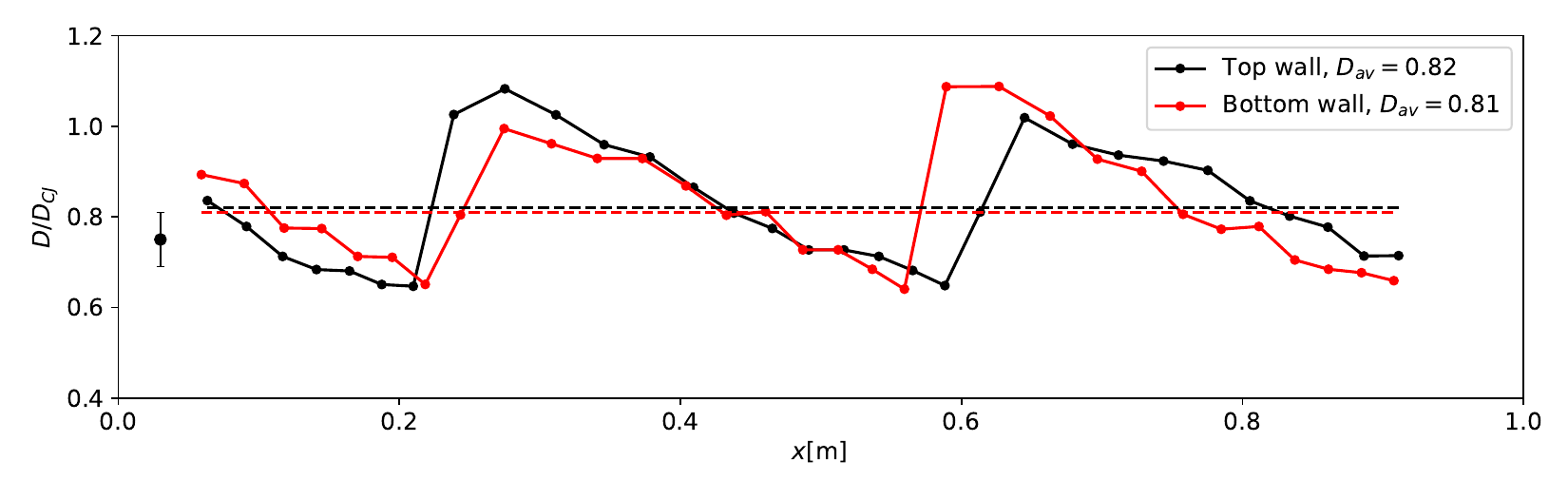}
        \caption{$\text{2H}_2/\text{O}_2/7\text{Ar}$ at $p_0 = 4.1$ \si{\kilo\pascal}}
        \label{fig:Ar-4-Dx-shad}
    \end{subfigure}
    \hfill
    \begin{subfigure}[b]{0.75\textwidth}
        \centering
        \includegraphics[width=\textwidth]{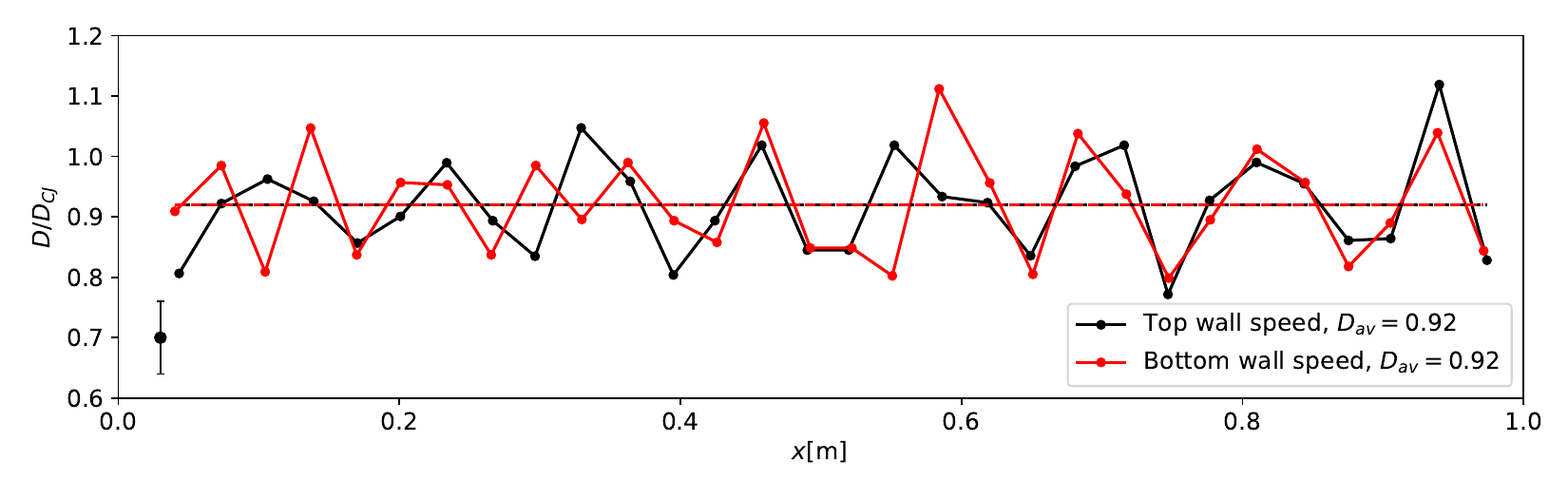}
        \caption{$\text{2H}_2/\text{O}_2/7\text{Ar}$ at $p_0 = 7.2$ \si{\kilo\pascal}}
        \label{fig:Ar-7-Dx-shad}
    \end{subfigure}
    \vfill
    \begin{subfigure}[b]{0.75\textwidth}
        \centering
        \includegraphics[width=\textwidth]{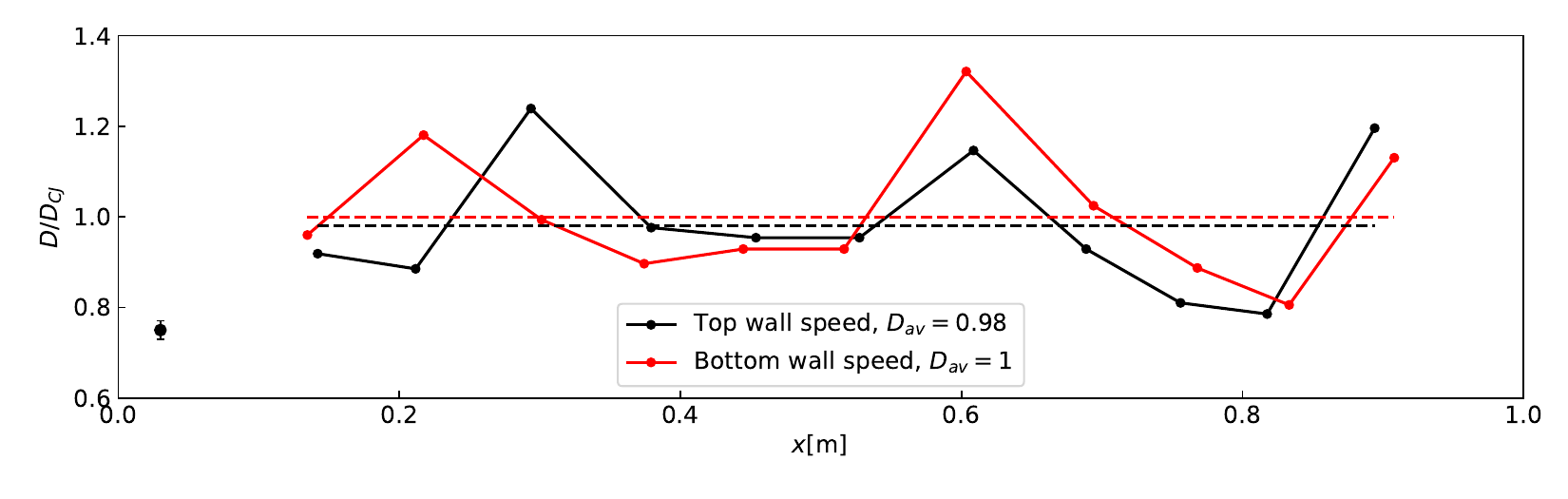}
        \caption{$\text{2H}_2/\text{O}_2/7\text{He}$ at $p_0 = 6.6$ \si{\kilo\pascal}}
        \label{fig:He-6-Dx-shad}
    \end{subfigure}
    \hfill
    \begin{subfigure}[b]{0.75\textwidth}
        \centering
        \includegraphics[width=\textwidth]{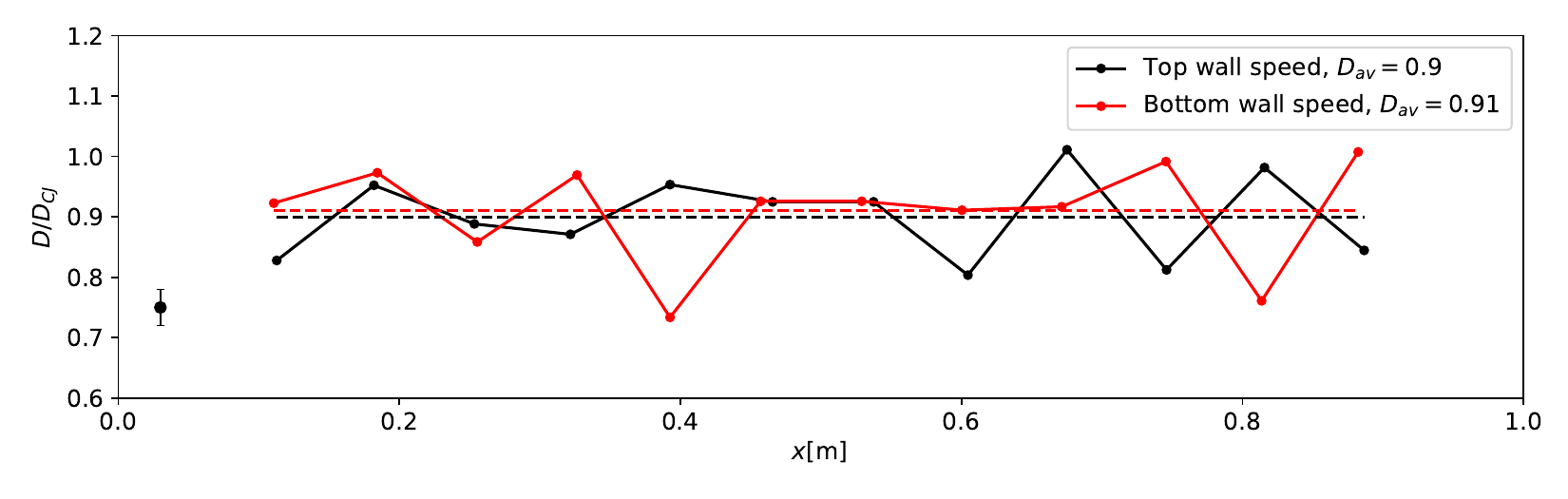}
        \caption{$\text{2H}_2/\text{O}_2/7\text{He}$ at $p_0 = 9.3$ \si{\kilo\pascal}}
        \label{fig:He-9-Dx-shad}
    \end{subfigure}
    \vfill
    \begin{subfigure}[b]{0.75\textwidth}
        \centering
        \includegraphics[width=\textwidth]{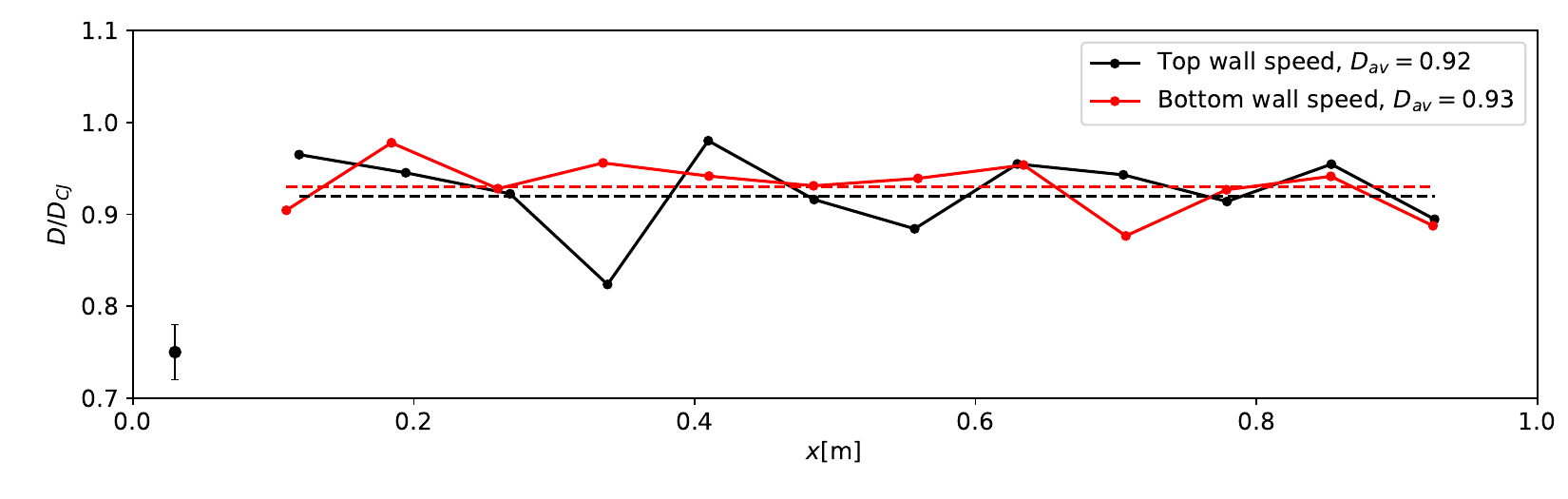}
        \caption{$\text{2H}_2/\text{O}_2/7\text{He}$ at $p_0 = 15$ \si{\kilo\pascal}}
        \label{fig:He-15-Dx-shad}
    \end{subfigure}
    \caption{Shock speed evolution over the length of the test section, extracted from shadowgraph images.. }
    \label{fig:shadVel}
\end{figure}
\begin{table*}[t!]
\centering
\caption{Detonation global propagation speed and cell width measured from the experiments; cell sizes marked with * are mode locked in a single cell configuration in our channel having the same dimension.}
\begin{tabular}{|c|c|c|c|}
\hline
 Mixture          & $p_0$ [\si{\kilo\pascal}] & $D/D_{CJ}$ & $\lambda$ [mm]   \\ 
\hline
\multirow{2}{*}{$\text{2H}_\text{2}/\text{O}_\text{2}/7\text{Ar}$} & 4.1   &  $0.84 \pm 0.016$ & 203*   \\ \cline{2-4} & 7.2   &  $0.9 \pm 0.01$ &$59\pm5$   \\ 
\hline
\multirow{3}{*}{$\text{2H}_\text{2}/\text{O}_\text{2}/7\text{He}$} & 6.6   &  $0.84\pm 0.014$  & 203*  \\ \cline{2-4} & 9.3   & $0.88 \pm 0.016$ &  $87 \pm 11$  \\ \cline{2-4}  & 15    &  $0.92 \pm 0.015$ &  $53 \pm 4$  \\ 
\hline
\end{tabular}
\label{tab:Exp}
\end{table*}

The global propagation speed, $D$, of the detonation wave over the 0.8 \si{\meter} distance between the first and last pressure sensors in the test section was determined using the time-of-arrival method, as outlined in Table \ref{tab:Exp}. The reported global velocity and its standard deviation represent the average values from repeated experiments under the same conditions (5 trials for argon-diluted mixtures and 10 trials for helium-diluted mixtures). The cell width, \(\lambda\), is the average measured from all frames of the schlieren and shadowgraph images across the repeated experiments, along with their standard deviation. For $\Delta_{i,CJ} = 2.8$ mm, the argon-diluted mixture exhibits a 16\% velocity deficit and a cell size of 0.203 \si{\meter}, while the helium-diluted mixture shows an 12\% velocity deficit with a cell size approximately half as large.

Figure \ref{fig:expArHeHigh} displays schlieren images of the detonation reaction zone structures at higher initial pressures for both mixtures, with the induction zone length kept constant at $\Delta_i = 1.6$ \si{\milli\meter}. By increasing the initial pressure, thereby enhancing the kinetic sensitivity of the mixture, the detonations exhibit considerably smaller cellular structures, with velocity deficits reduced to 10\% for the argon-diluted mixture and 8\% for the helium-diluted mixture. Although the cell size is slightly smaller in the helium-diluted case, the overall regular cellular structure of the detonations appears similar between the two mixtures. Additionally, possible 3D-like effects in the detonation structure can be observed in both mixtures, as indicated by the presence of duplicate features that do not overlap in the schlieren images.

The locally averaged speeds along both the top and bottom walls were calculated from the shadowgraph images in each experiment, with the distance between every two consecutive frames divided by their time interval, $\Delta t$. The resulting speed profiles are shown in Fig.\ \ref{fig:shadVel}, normalized by the ideal CJ detonation velocity. For the argon-diluted mixture at $p_0 = 4.1$ \si{\kilo\pascal} (Fig.\ \ref{fig:Ar-4-Dx-shad}), the velocity decays within a single cell from 1.1 to 0.6 and repeats throughout the test section. When the pressure is increased to 7.2 \si{\kilo\pascal} (Fig.\ \ref{fig:Ar-7-Dx-shad}), the average velocity becomes 0.92 $D_{CJ}$. The overall trends suggest that the detonation propagation speed remains nearly steady at a macro scale within the test section. A similar quasi-steady propagation is observed in the helium-diluted mixture; however, due to the higher propagation speed in this mixture and limitations in the frame rate of the shadowgraph imaging, fewer data points are available (Fig.\ \ref{fig:He-6-Dx-shad}, \ref{fig:He-9-Dx-shad}, \ref{fig:He-15-Dx-shad}).
\begin{figure}[t!]
    \centering
    \begin{subfigure}[b]{0.45\textwidth}
        \centering
        \includegraphics[width=\textwidth]{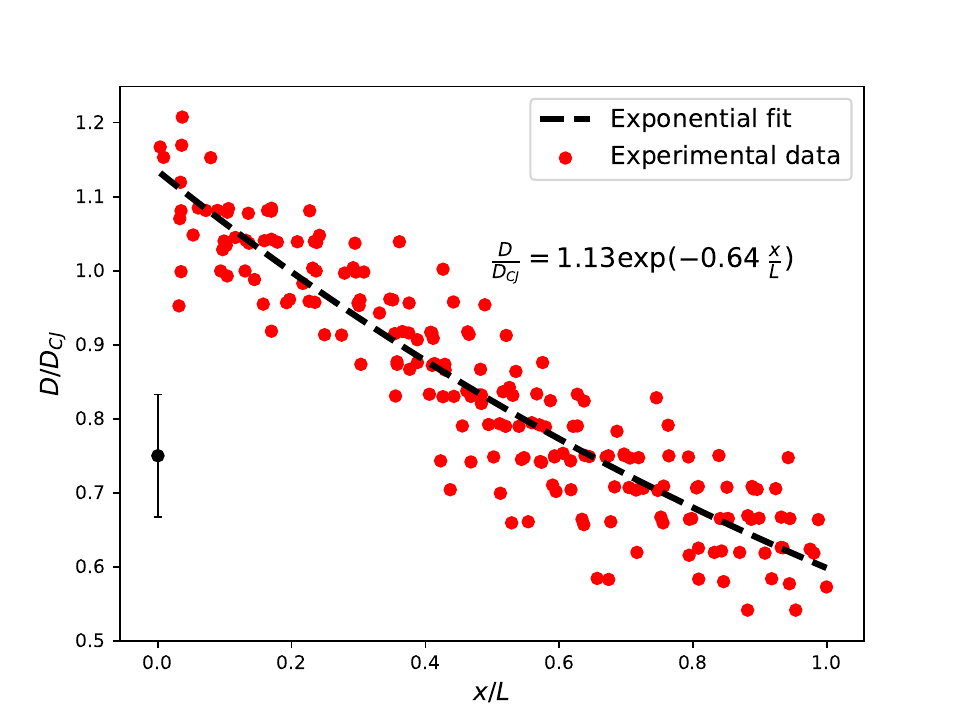}
        \caption{$\text{2H}_2/\text{O}_2/7\text{Ar}$ at $p_0 = 4.1$ \si{\kilo\pascal}}
        \label{fig:Ar-cell4}
    \end{subfigure}
    \hfill
    \begin{subfigure}[b]{0.45\textwidth}
        \centering
        \includegraphics[width=\textwidth]{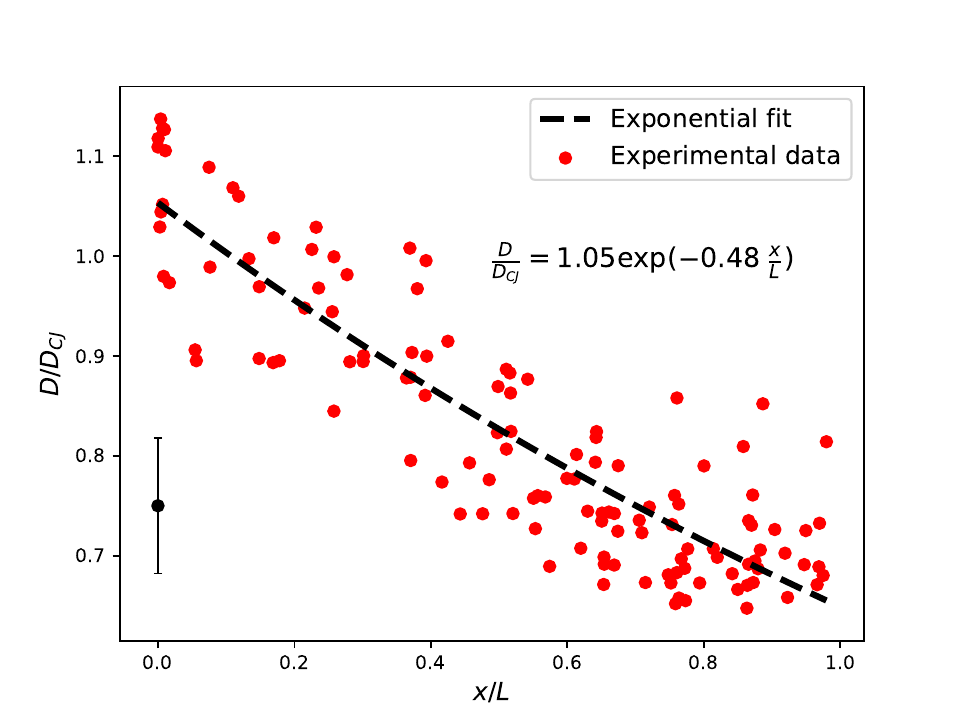}
        \caption{$\text{2H}_2/\text{O}_2/7\text{He}$ at $p_0 = 6.6$ \si{\kilo\pascal}}
        \label{fig:He-cell6}
    \end{subfigure}
    \vfill
    \begin{subfigure}[b]{0.45\textwidth}
        \centering
        \includegraphics[width=\textwidth]{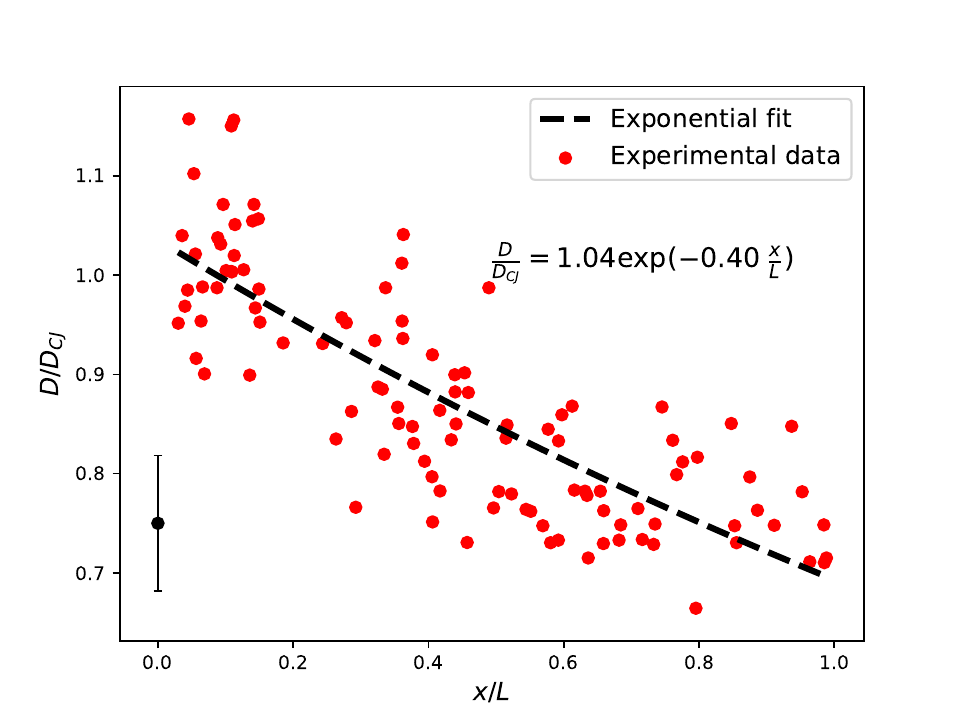}
        \caption{$\text{2H}_2/\text{O}_2/7\text{He}$ at $p_0 = 9.3$ \si{\kilo\pascal}}
        \label{fig:He-cell15}
    \end{subfigure}
    \caption{Shock speed evolution over a detonation cell, with the exponential curve fit shown in black. The experimental data were obtained from repeated experiments conducted \ref{fig:Ar-cell4}) 5 times, \ref{fig:He-cell6}) and \ref{fig:He-cell15}) 10 times. }
    \label{fig:exponecell}
\end{figure}
We further reconstructed the velocity profile within a single cellular cycle for experiments with sufficiently large cells, following the methodology detailed in Cheevers' thesis \cite{cheevers2021optical}. For the $\text{2H}_2/\text{O}_2/7\text{Ar}$ mixture at $p_0 = 4.1$ \si{\kilo\pascal}, data were obtained from a single cell across five repeated experiments. For the $\text{2H}_2/\text{O}_2/7\text{He}$ mixture, data were collected from ten repeated experiments at each pressure (6.6 \si{\kilo\pascal} and 9.3 \si{\kilo\pascal}). An exponential curve was fitted to the data using the equation \( \frac{D}{D_{CJ}} = \frac{D_0}{D_{CJ}} \exp\left(-\frac{bx}{L}\right) \), where \( D_0 \) represents the velocity at the start of the cell, \( L \) is the cell length, and \( b \) is the decay rate. These results are illustrated in Fig.\ \ref{fig:exponecell}.  The highest decay rate within a single cell was observed in the $\text{2H}_2/\text{O}_2/7\text{Ar}$ mixture, where the velocity starts at 1.13 at the beginning of the cell and decays to 0.5 by the end. In the $\text{2H}_2/\text{O}_2/7\text{He}$ mixture, the velocity decreases from 1.05 to 0.65 at $p_0 = 6.6$ \si{\kilo\pascal} and from 1.04 to 0.7 at $p_0 = 9.3$ \si{\kilo\pascal}.

Overall, the experimental measurements of cell width reveal a significant discrepancy in cell size between the two mixtures diluted with argon and helium, despite having the same induction zone length at lower pressures. This raises the question: Is the discrepancy a result of differences in non-equilibrium effects, or is it due to boundary layer losses present in the experiments? In the following sections, we will explore this question by estimating the characteristic vibrational relaxation times and conducting boundary layer loss calculations.

\section{Vibrational non-equilibrium effect}
\label{sec:Vib}
Recent modelling efforts on non-equilibrium effects in detonations suggest that hydrogen-based detonations may be significantly influenced by the vibrational non-equilibrium of molecules, as vibrational modes are the last to reach equilibrium \cite{taylor2013estimates, taylor2013numerical}. This phenomenon could help explain the discrepancies observed in experimentally measured cell sizes, as differences in molecular weight and reduced mass of the bath gases impact molecular collisions and alter relaxation rates \cite{shi2017assessment}. In this section, we aim to estimate the vibrational relaxation time of the \( \text{H}_2 \) and \( \text{O}_2 \) molecules in the two mixtures studied. By using available empirical data, we will calculate the vibrational relaxation time and compare it to the ignition delay times. This comparison will help determine if the discrepancies in the cellular structure of detonation observed in the previous section are due to differences in vibrational non-equilibrium effects of diluents.
\begin{table*}[t]
\centering
\caption{Vibrational relaxation time and ignition delay time calculations.}
\resizebox{\textwidth}{!}{
\begin{tabular}{|c|c|c|c|c|c|c|}
\hline
 Mixture & $p_0$ [\si{\kilo\pascal}] & $\tau_{H_2}$ [s] & $\tau_{O_2}$ [s] & $\tau_{\text{ignition}}$ [s] & $\frac{\tau_{\text{ignition}}}{\tau_{H_2}}$ & $\frac{\tau_{\text{ignition}}}{\tau_{O_2}}$ \\
\hline
\multirow{2}{*}{ $\text{2H}_2/\text{O}_2/7\text{Ar}$} & 4.1 & $3.05 \times 10^{-6}$ & $8.4 \times 10^{-7}$ & $7.03 \times 10^{-6}$ & 2.3 & 8.4 \\ \cline{2-7}
& 7.2 & $1.63 \times 10^{-6}$ & $4.6 \times 10^{-7}$ & $4 \times 10^{-6}$ & 2.4 & 8.7 \\
\hline
\multirow{2}{*}{ $\text{2H}_2/\text{O}_2/7\text{He}$} & 9.3 & $8.1 \times 10^{-7}$ & $1.6 \times 10^{-7}$ & $3.04 \times 10^{-6}$ & 3.7 & 18.5 \\ \cline{2-7}
& 15 & $4.8 \times 10^{-7}$ & $9.7 \times 10^{-8}$ & $1.8 \times 10^{-6}$ & 3.7 & 18.2 \\
\hline
\end{tabular}}
\label{tab:VibIgn}
\end{table*}
The vibrational relaxation time of each species, \( \tau_i \), in a mixture of \( N \) gases can be computed using Eq.\ \eqref{eq:relaxation_time}. This equation takes into account the contributions of collisions with the species itself and other species, based on fitting experimental data using a least squares method \cite{dove1974vibrational}.
\begin{equation}
\frac{1}{p\tau_i} = \sum_{j=1}^{N} \frac{X_j}{p\tau_{i-j}}
 \label{eq:relaxation_time}
\end{equation}
where $p$ is the pressure, $\tau_i$ is the vibrational relaxation time of species $i$, $X_j$ is the mole fraction of species $j$ in the mixture and $\tau_{i-j}$ is the vibrational relaxation time of the interaction between species $i$ and $j$. Therefore, the vibrational relaxation time of the hydrogen and oxygen can be estimated as follows:
\begin{equation}
\frac{1}{p\tau_{H_2}} = \frac{X_{H_2}}{p\tau_{H_2-H_2}} + \frac{X_{O_2}}{p\tau_{H_2-O_2}} + \frac{X_{\text{diluent}}}{p\tau_{H_2-\text{diluent}}}
\end{equation}
\begin{equation}
\frac{1}{p\tau_{O_2}} = \frac{X_{H_2}}{p\tau_{O_2-H_2}} + \frac{X_{O_2}}{p\tau_{O_2-O_2}} + \frac{X_{\text{diluent}}}{p\tau_{O_2-\text{diluent}}}
\end{equation}
In this study, $X_{H_2} = 0.2$, $X_{O_2} = 0.1$, and $X_{\text{diluent}} = 0.7$. To calculate $p\tau_{H_2-j}$, we use the following empirical relations \cite{dove1974vibrational}:
\begin{equation}
\log{p\tau_{H_2-j}} = A T^{-1/3} - B
\end{equation}
where \( T \) and \( p \) represent the post-shock temperature and pressure, respectively. The parameter \( A \) takes the values 34.74, 41.35, and 45.09, while the parameter \( B \) is 8.686, 8.984, and 8.956 for \(\text{H}_2\)-\(\text{H}_2\), \(\text{H}_2\)-He, and \(\text{H}_2\)-Ar interactions, respectively. Due to the unavailability of experimental data for the relaxation time of $\text{H}_2$-$\text{O}_2$, we employed the relaxation time of $\text{H}_2$-$\text{Ar}$ as a substitute. This approach is justified by the similarity in molecular weights between argon (40 AMU) and oxygen ($\text{O}_2$: 32 AMU), as suggested by Taylor et al.\cite{taylor2013estimates}. For the relaxation time of the oxygen $p\tau_{O_2-j}$, we use the following equation in which parameters $A$ and $B$ are found empirically \cite{millikan1963systematics, taylor2013estimates}.
\begin{equation}
p\tau_{O_2-j}= \exp\left[C\left(T^{-1/3} - D\right) - 18.42\right]
\end{equation}
\begin{figure}[t]
    \centering
    \begin{subfigure}[b]{0.45\textwidth}
        \centering
	\includegraphics[width=\textwidth]{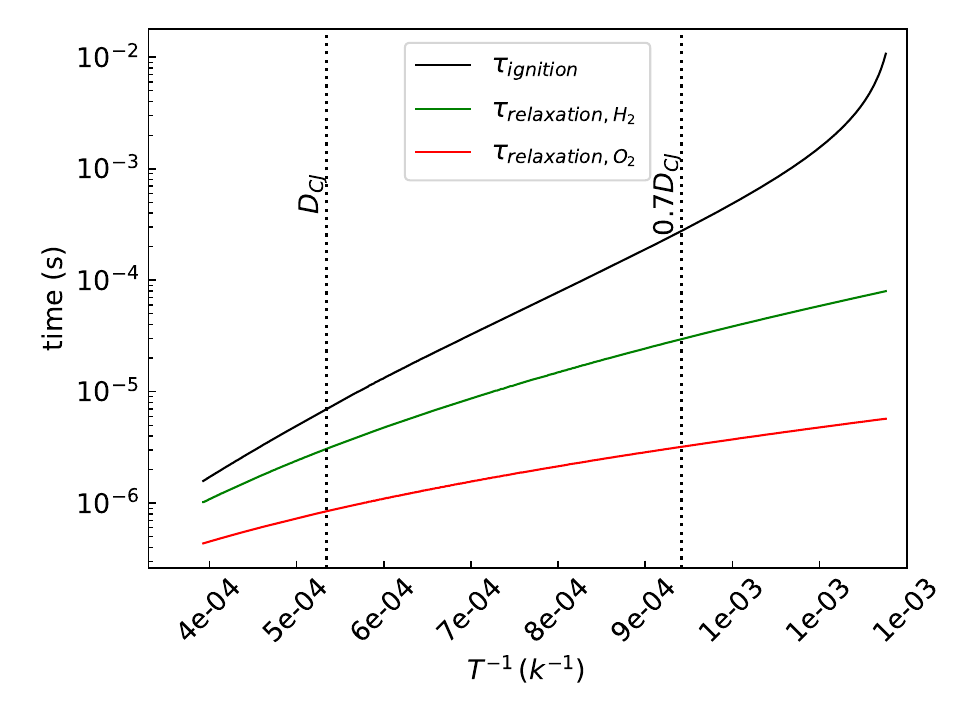}
        \caption{$\text{2H}_2/\text{O}_2/7\text{Ar}$ at $p_0 = 4.1$ \si{\kilo\pascal}}
        \label{fig:relaxAr}
    \end{subfigure}
    \hfill
    \begin{subfigure}[b]{0.45\textwidth}
        \centering
	\includegraphics[width=\textwidth]{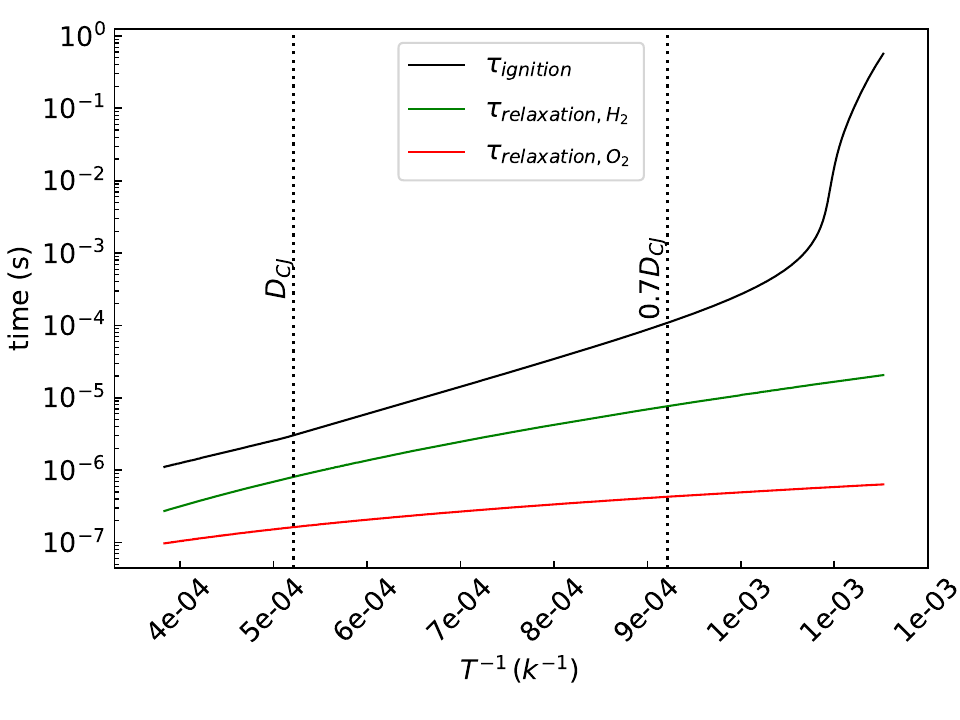}
        \caption{$\text{2H}_2/\text{O}_2/7\text{He}$ at $p_0 = 9.3$\si{\kilo\pascal}}
        \label{fig:relaxHe}
    \end{subfigure}
	\caption{Comparisons of the ignition delay time and vibrational relaxation time versus the post-shock temperature.}
        \label{fig:tigtrelax}
\end{figure}

\noindent where \( C = 133 \) and \( D = 0.03 \) for \( \text{O}_2\)-\( \text{O}_2 \), and \( C = 36 \) and \(  D = 0.000067 \) for \( \text{O}_2\)-\( \text{H}_2 \). For the relaxation time of oxygen with diluents (\( \text{O}_2\)-Ar and \( \text{O}_2\)-He), we use the experimental values provided by Millikan and White, where \( p\tau_{\text{O}_2-\text{Ar}} = 0.000038 \ \text{atm.s} \) and \( p\tau_{\text{O}_2-\text{Ar}} = 0.000000509  \ \text{atm.s}\) respectively \cite{millikan1963systematics}.\\

Table \ref{tab:VibIgn} presents the results of the vibrational relaxation time scales for the mixtures of  $\text{2H}_2/\text{O}_2/7\text{Ar}$ and  $\text{2H}_2/\text{O}_2/7\text{He}$ at initial pressures of the experiments computed at the Von Neumann state of the detonations.  Additionally, the ignition delay time, computed using a realistic chemistry ZND calculation for detonation waves in the two mixtures, is also shown to provide a qualitative comparison. 

The results show that helium as a diluent reduces the relaxation times of hydrogen and oxygen by approximately 70\% as compared to argon.  The ratios between ignition delay time and vibrational relaxation time show that in all cases, relaxation is faster than the induction kinetics.  The slowest relaxation is that of H$_2$ in the argon diluted system, which is approximately half the induction time predicted by the ZND model.  This may suggest the non-negligible effect of vibrational relaxation, as argued by Taylor et al.\ \cite{taylor2013numerical}.  

We further calculated how these time scales vary for different shock speeds, since the cellular dynamics are transient.  The ignition delay time in the post-shock region was calculated using constant-volume homogeneous reactor of Cantera thermochemical tools \cite{cantera} and the San Diego mechanism \cite{sandiego}. The results are shown in Figure \ref{fig:tigtrelax}.  The temperature dependence of induction kinetics being different, the relaxation rates become an order of magnitude slower than the induction kinetics at the low speed stages of the detonation structure. Owing to the lower sensitivity of relaxation rates, calculations of induction delays along particle paths accounting for the expansion associated with shock non-steadiness would further exacerbate this discrepancy \cite{cheevers2021optical}.  It can thus be argued that induction become much larger than relaxation times once their sensitivity to flow non-steadiness prevalent in cellular dynamics is incorporated.     


For the same induction zone length, we find that the helium diluted system is more reactive.  It has a lower velocity deficit and smaller cells than in the argon system.  This trend cannot be reconciled by the vibrational relaxation explanation.  The relaxation is slower with argon, and hence one would expect at larger translational temperature in the induction zone and a speed up in reactivity.  This is not the case in the experiments, where the opposite trend is observed.

\section{Numerical modelling of boundary layer losses}
\label{sec:Num}
\subsection{Model formulation}
\begin{figure}[t]
\centering
\begin{subfigure}{0.45\textwidth}
    \includegraphics[scale=0.3]{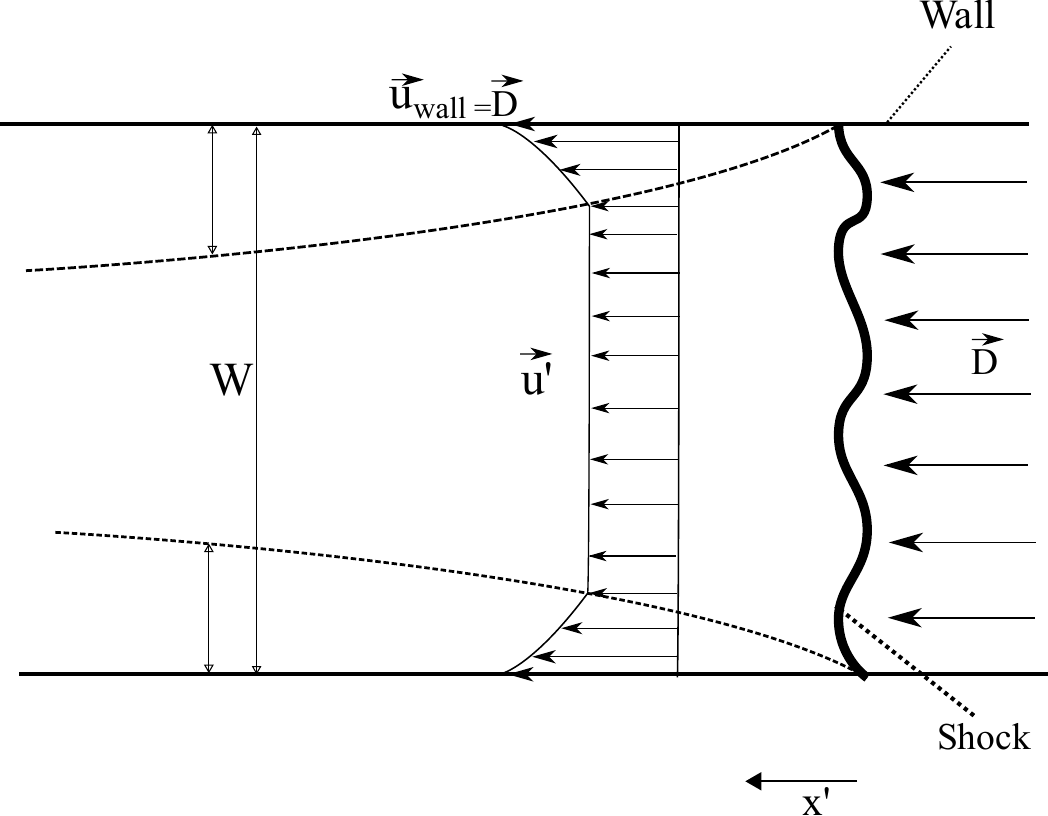}
    \caption{}
    \label{fig:bl-a}
\end{subfigure}
\hfill
\begin{subfigure}{0.45\textwidth}
    \includegraphics[scale=0.3]{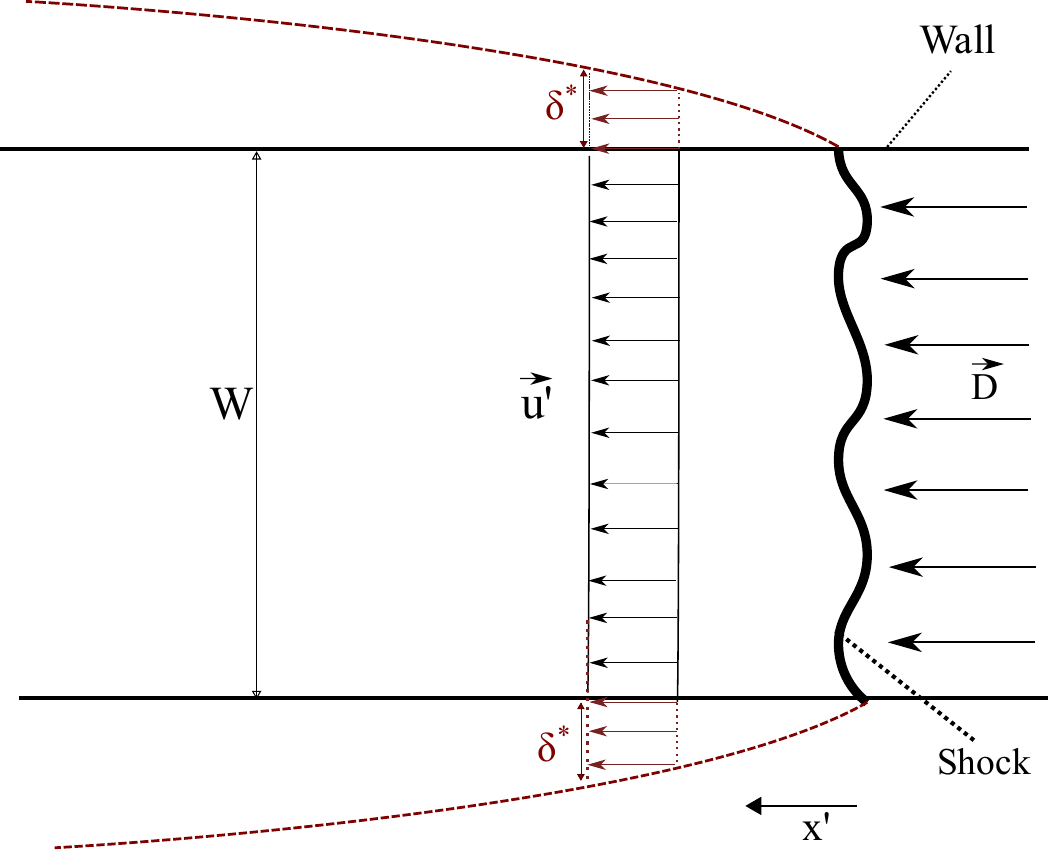}
    \caption{ } 
\label{fig:bl-b}
\end{subfigure} 
\caption{The steady flow in the post-shock region of a detonation wave in the shock-attached reference frame. The dashed lines inside the walls (a) indicate the outer edges of the boundary layer ($\delta$) and the dashed lines outside of the walls (b) indicate the outer edges of the displaced boundary layer ($\delta^\star$). $W$ is the channel thickness in the z-direction, $D$ is the detonation speed, $x'$ is the distance in the post-shock region in the shock-attached frame and $u'$ is the post-shock region velocity.}
\label{fig:blMirel}
\end{figure}

We proceeded by establishing if the results of our experiments, namely anomalous differences in cell sizes and velocity deficits for the same induction zone length,  are compatible with boundary layer losses.  To investigate this hypothesis, we employ the method proposed by Xiao et al. \cite{Xiao:2021}, who introduced a quasi-2D model to reconstruct the three-dimensional cellular structures of detonation waves in narrow channels. Their model incorporates an area divergence term into the reactive Euler equations to account for boundary layer losses in the third dimension, as illustrated in Fig.\ \ref{fig:blMirel}. Figure \ref{fig:bl-a} shows the velocity gradients near the walls caused by the presence of a boundary layer in contrast to the uniform flow observed in the core region. Figure \ref{fig:bl-b} depicts an equivalent approach in which the effect of the boundary layer losses are accounted for by incorporating the flow divergence in the third dimension due to the negative boundary layer displacement thickness, $\delta^{*}$. The area divergence term is evaluated based on Mirels’ laminar boundary layer solution \cite{Mirels:1956}.  An improvement on the account for flow acceleration is the only difference from the treatment used by \cite{Xiao:2021}.  This was first proposed by the authors in our preliminary work \cite{ZangeneIcder:2022} and very recently further validated by Smith et al. \cite{smith2024nature} using a higher fidelity chemical model. 

We briefly review the model formulation - for details, see \cite{Xiao:2021}.  The governing equations are the reactive two-dimensional Euler equations supplemented by a source term to model the boundary layer effects, outlined as follows:
\begin{equation}
\frac{\partial \rho}{\partial t} + \frac{\partial (\rho u)}{\partial x} + \frac{\partial (\rho v)}{\partial y} = -\rho \frac{1}{A} \frac{DA}{Dt} 
\label{eq:mass}
\end{equation}
\begin{equation}
\frac{\partial \rho u}{\partial t} + \frac{\partial (\rho u^{2}+p)}{\partial x} + \frac{\partial (\rho uv)}{\partial y} = -\rho u \frac{1}{A} \frac{DA}{Dt} 
\label{eq:momX}
\end{equation}
\begin{equation}
\frac{\partial \rho v}{\partial t} + \frac{\partial (\rho uv)}{\partial x}+ \frac{\partial (\rho v^{2}+p)}{\partial y}  = -\rho v \frac{1}{A} \frac{DA}{Dt} 
\label{eq:momY}
\end{equation}
\begin{equation}
\frac{\partial (\rho e)}{\partial t} + \frac{\partial (\rho eu+pu)}{\partial x} + \frac{\partial (\rho ev+pv)}{\partial y} = -(\rho e + p) \frac{1}{A} \frac{DA}{Dt} -Q\dot\omega_{R}
\label{eq:energy}
\end{equation}
\begin{equation}
\frac{\partial (\rho Y)}{\partial t} + \frac{\partial (\rho uY)}{\partial x} + \frac{\partial (\rho vY)}{\partial y} = -\rho Y \frac{1}{A} \frac{DA}{Dt} -\dot\omega_{R}
\label{eq:sp}
\end{equation}
where $\rho$, u, v, A, p, Q, Y, and $\dot\omega_{R}$ represent the density of the mixture, velocity in the x-direction, velocity in the y-direction, cross-sectional area, pressure, heat release, mass fraction, and the rate of mass production of the individual reactant R. The total energy is defined as the sum of the sensible energy plus the kinetic energy,  $e = \frac{p/\rho}{\gamma - 1} + \frac{1}{2}(u^2 + v^2)$ where $\gamma$ is the ratio of specific heats. 

The viscous boundary layer developing behind a detonation wave leads to an apparent cross-sectional increase to $A=W+2\delta^*$ due to the negative mass displacement thickness $\delta^*$, as illustrated in Fig.\ \ref{fig:blMirel}.  The rate of lateral strain of a fluid element is thus 
\begin{equation}
\frac{D(\ln A)}{Dt}\cong\frac{2}{W}\frac{D \delta^*}{Dt}
\label{eq:1}
\end{equation}
A special treatment is required for accounting for the boundary layer growth and its negative displacement thickness in non-uniform flow.  In the absolute frame of reference of the calculations, the flow in the reaction zone decelerates strongly while it expands during the exothermic events.  To address this, we model the boundary layer growth in terms of the time elapsed since having crossed the shock, since $\delta \sim \sqrt{t_{elapsed}}$ will continue to offer a good approximation for boundary layer growth even for non steady or steady accelerating flow. We also assume the displacement thickness proportional to the boundary layer thickness, a reasonable assumption.   This amounts to generalizing the expression for the negative displacement thickness found by Mirels \cite{Mirels:1956} for laminar flow from 
\begin{equation}
\delta^* = K_M \sqrt{\nu \frac{x'}{u'}}
\label{eq:deltaStarorgin}
\end{equation}
to 
\begin{equation}
\delta^* = K_M \sqrt{\nu \int_{0}^{x'} \frac{dx'}{u'(x')}}= K_M \sqrt{\nu t_{elapsed}}
\label{eq:deltaStar}
\end{equation}

\noindent in which $W$ is the physical channel width of 19 \si{\milli\meter}. Mirels' constant $K_M$ can be calculated separately, as for steady flow, and depends on the mixture composition and post shock state. The detailed computation of this constant is available in the work of Xiao and Radulescu \cite{xiao:2020dynamics}.  In the case of $\text{2H}_\text{2}/\text{O}_\text{2}/7\text{Ar}$ mixture, the determined value of $K_M$ is approximately 4, while for the mixture of $\text{2H}_\text{2}/\text{O}_\text{2}/7\text{He}$, the value of $K_M$ is found to be 3.8.  

Using \eqref{eq:deltaStar}, the lateral strain rate required in the computations is 
\begin{equation}
\frac{D(\ln A)}{Dt}\cong\frac{K_M}{W}\sqrt{\frac{\nu}{t_{elapsed}}}
\label{eq:source}
\end{equation}
This simple correction recovers approximately the boundary layer growth in steady flow with non-zero pressure gradients.  An observer following an accelerating flow due to a local favourable pressure gradient sees a smaller boundary layer than an observer having reached the same position having propagated at a constant speed.   This is due to the time elapsed since crossing the shock being smaller for the observer that reached that location faster.  

Since the flow is non-steady, the time elapsed is a property of a given fluid particle, i.e., the time elapsed since the wall started pulling away from that particle, as in Rayleigh's suddenly accelerating plate problem.  The only information required in the calculation is thus the time at which a fluid particle has crossed the shock $t_s$, since $t_{elapsed} = t - t_s$.  Since each particle retains its value of $t_s$, this is updated only once, when passing a shock, and then evolved as a passive scalar advected with the flow, satisfying: 

\begin{equation}
\frac{\partial t_s}{\partial t}+ u\frac{\partial t_s}{\partial x}+v\frac{\partial t_s}{\partial y}=0
\end{equation}

The same model was suggested first by Xiao et al. \cite{Xiao:2021} in their write-up.  Unfortunately, their calculations used a non-advected evolution of $t_s$.  The incorrect write-up was not noted at the time.  As we will show below, the presently suggested advected form does not require the artificial re-tuning of the Mirels' constant, as performed by Xiao et al.\ \cite{Xiao:2021} in order to match experiment.  This was also communicated subsequent to our first dissemination of our finding \cite{ZangeneIcder:2022} by Smith et al.\ \cite{smith2024nature}, who also found it to be better approximation. 

\begin{figure}[t!]
    \centering
    \begin{subfigure}[b]{0.45\textwidth}
        \centering
	\includegraphics[scale=0.34]{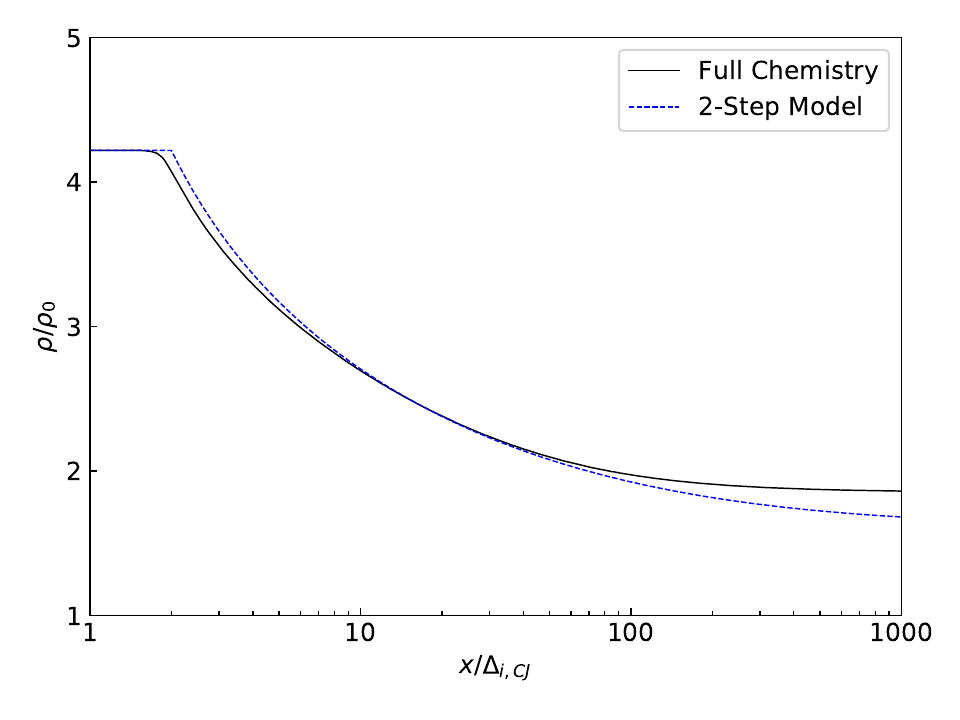}
        \caption{$\text{2H}_2/\text{O}_2/7\text{Ar}$ at $p_0 = 4.1$ \si{\kilo\pascal}}
        \label{fig:Ar-cell4-chem}
    \end{subfigure}
    \hfill
    \begin{subfigure}[b]{0.45\textwidth}
        \centering
	\includegraphics[scale=0.34]{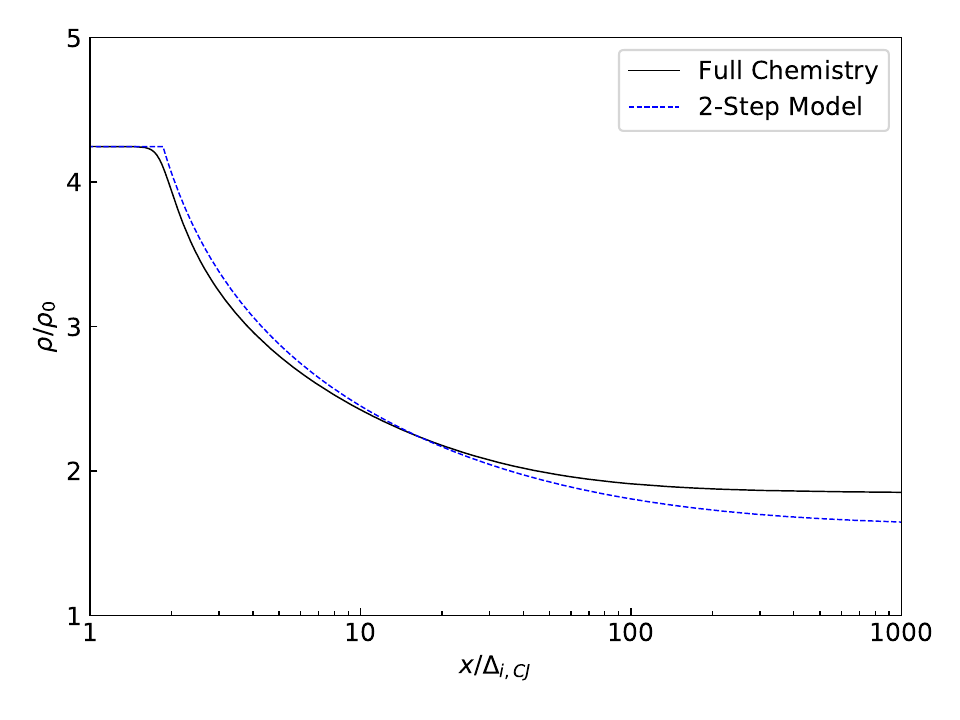}
        \caption{$\text{2H}_2/\text{O}_2/7\text{Ar}$ at $p_0 = 9.3$ \si{\kilo\pascal}}
        \label{fig:He-cell6-chem}
    \end{subfigure}
	\caption{Calibration of the two-step model based on detailed chemistry. $\rho$ represents the post-shock density, $\rho_{0}$ is the initial density, $x$ denotes the distance, and $\Delta_{i, CJ}$ is the induction zone length.}
	\label{fig:calib}
\end{figure}
\begin{table}[t!]
\small
  \centering
      \caption{Calibrated non-dimensional parameters for the two-step model from the detailed chemistry.}
    \begin{tabular}{|l|c|c|c|c|c|c|c|c|c|} 
\hline
  Mixture & $p_0$ [\si{\kilo\pascal}] & $\gamma$ & $E_a/RT_0$ & $Q/RT_0$ & $k_i$ & $k_r$ &$\alpha$ &$\beta$ &$\nu$  \\
      \hline
$\text{2H}_2/\text{O}_2/7\text{Ar}$&  $4.1$  & $1.5$ & $31.1$ & $11.5$ & $45.5$ & $0.08$ &1.2 &1 &1.6 \\
      \hline
$\text{2H}_2/\text{O}_2/7\text{He}$ &   $9.3$ & $1.5$ & $23.9$ & $11.9$ & $12.2$ & $0.13$ &1.2 &1 &1.5 \\
      \hline
    \end{tabular}
 \label{tab:chemist}
\end{table}
To model the chemical kinetics, the two-step chain-branching reaction model (the thermally neutral induction zone followed by the exothermic reaction zone) was employed and the non-dimensional parameters for the two-step model were calibrated from the detailed chemistry by using the San Diego chemical reaction mechanism \cite{sandiego} and Shock and Detonation Toolbox (SDToolbox) \cite{SDToolbox}, shown in Fig.\ \ref{fig:calib}. The induction and reaction transport equations can be formulated as Eq.\ \eqref{eq:chemistI} and \eqref{eq:chemistR}.
\begin{equation}
\frac{\partial (\rho \lambda_{i})}{\partial t} + \frac{\partial (\rho u \lambda_{i})}{\partial x} + \frac{\partial (\rho v \lambda_{i})}{\partial y} = -\rho \lambda_{i} \frac{1}{A} \frac{DA}{Dt} -\mathcal{H}(\lambda_{i})k_{i} \rho ^{\alpha+1}exp(-\frac{Ea}{RT})
\label{eq:chemistI}
\end{equation}
\begin{equation}
\frac{\partial (\rho \lambda_{r})}{\partial t} + \frac{\partial (\rho u \lambda_{r})}{\partial x} + {\partial (\rho v \lambda_{r})}{\partial y} = -\rho \lambda_{r} \frac{1}{A} \frac{DA}{Dt} [1-\mathcal{H}(\lambda_{i})]k_{r} \rho ^{\beta+1}\lambda_{r}^{\nu}
\label{eq:chemistR}
\end{equation}
\begin{figure*}[t]
\centering
\includegraphics[scale=0.4]{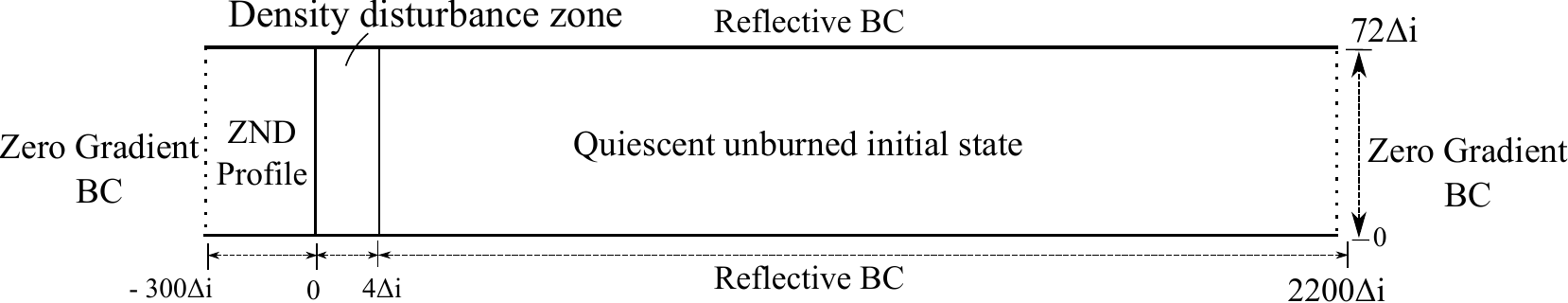}
\caption{Initial and boundary conditions for the 2D detonation propagation simulation, where 72$\Delta_{i,CJ}$ corresponds to the actual channel height of 203 \si{\milli\meter} in the experiment.}
\label{fig:BC}
\end{figure*}
The progress variable for the induction zone, $\lambda_{i}$, has a value of 1 ($\mathcal{H}(\lambda_{i})=1$) at the reactants and 0 ($\mathcal{H}(\lambda_{i})=0$) at the induction zone's termination. Correspondingly, the reaction progress variable, $\lambda_{r}$, holds a value of 1 in the unburned zone and 0 in the burned products. $k_i$ and $k_r$ represent rate constants, $\nu$ represents the reaction order, and $\alpha$ and $\beta$ stand as additional empirical reaction order parameters. The values of $\beta$, $\alpha$ and $\nu$ were chosen to align with the detailed chemistry ZND structure, which is readily discernible from the density profile illustrated in Fig.\ \ref{fig:calib}.

Table \ref{tab:chemist} shows the non-dimensional parameters for the two-step chemistry model for the two different mixtures. The post-shock isentropic exponent of the CJ detonation was represented by $\gamma$, and the heat release $Q$ was obtained using the perfect gas relation to ensure accurate recovery of the Mach number \cite{Fickett:1979}. The symbol $E_a$ represents the activation energy that impacts the temperature-dependent behaviour of the induction zone duration. It was calculated using the logarithmic derivative of the ignition delay time with respect to the inverse of post-shock temperature.

In this investigation, the normalization scales for the analysis include the initial state variables ($p_{0}$, $\rho_{0}$) and the ideal ZND induction zone length, $\Delta_{i,CJ}$.
\begin{figure}[t]
    \centering
        \includegraphics[width=\textwidth]{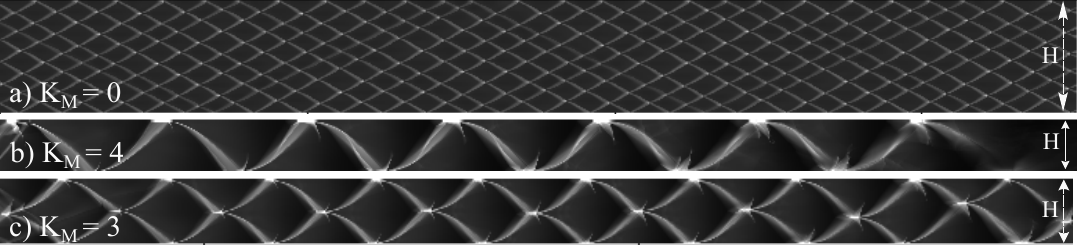}
    \caption{The recorded maximum energy release rates of detonations in the mixture of $\text{2H}_2/\text{O}2/7\text{Ar}$ at $p_0 = 4.1 $ \si{\kilo\pascal} and $T_0 = 293$ K.  The cell size and mean propagation speeds are: (a) \( D/D_{\text{CJ}} = 1 \), \( \lambda = 37 \, \text{mm} \), (b) \( D/D_{\text{CJ}} = 0.77 \), \( \lambda = 406 \, \text{mm} \), (c) \( D/D_{\text{CJ}} = 0.83 \), \( \lambda = 203 \, \text{mm} \). The channel height $H$ is 72 $\Delta_{i, CJ}$, corresponding to a shock tube channel height of 203 \si{\milli\meter}. }
    \label{fig:sootAr}
\end{figure}
\begin{figure}[t]
    \centering
        \includegraphics[width=\textwidth]{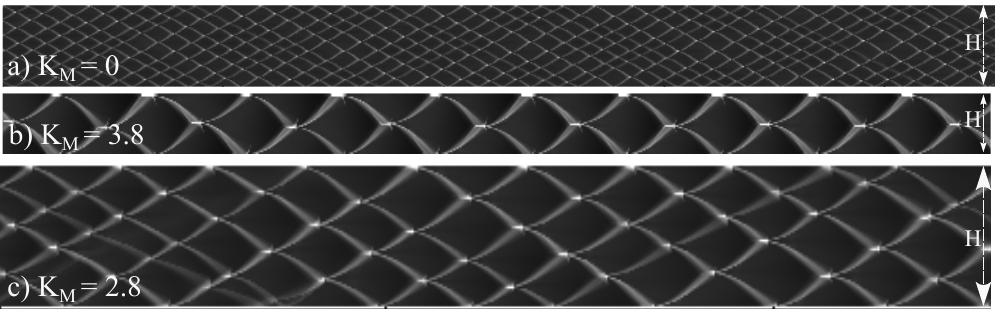}
    \caption{The recorded maximum energy release rates of detonations in the mixture of $\text{2H}_2/\text{O}2/7\text{He}$ at $p_0 = 9.3 $ \si{\kilo\pascal} and $T_0 = 293$ K. The cell size and mean propagation speeds are: (a) \( D/D_{\text{CJ}} = 1 \), \( \lambda = 37 \, \text{mm} \), (b) \( D/D_{\text{CJ}} = 0.81 \), \( \lambda = 203 \, \text{mm} \), (c) \( D/D_{\text{CJ}} = 0.88 \), \( \lambda = 81 \, \text{mm} \). The channel height $H$ is 72 $\Delta_{i, CJ}$, corresponding to a shock tube channel height of 203 \si{\milli\meter}. }
    \label{fig:sootHe}
\end{figure}
A second-order-accurate exact Godunov solver \cite{Falle:1991} with adaptive mesh refinement \cite{Xiao:2021} is used to solve the governing equations.  The induction zone length is kept constant for both helium and argon experiments. For reference, the induction zone length is 0.0028 m. The channel height is non-dimensionalized by the induction zone length. To make sure that we cover the length of the shock tube, 3.4 m, a domain length of 2500$\Delta_{i,CJ}$ is considered for the simulations.  In these simulations, the detonation propagates from left to right, with reflective boundary conditions imposed on the top and bottom sides, and zero-gradient boundary conditions applied to the left and right ends, illustrated in Fig.\ \ref{fig:BC}.

For the numerical resolution, the coarsest grid size was set to $1/2 \, \Delta_{i,CJ}$, and the finest grid size was set to $1/16 \, \Delta_{i,CJ}$. Since we use a two-step chemistry model similar to that employed by Xiao \cite{xiao2020thesis} the 5 levels of mesh refinement chosen in this study have been verified as sufficient.
\begin{figure}[t!]
    \centering
    \begin{subfigure}[b]{\textwidth}
        \centering
	\includegraphics[width=0.9\linewidth]{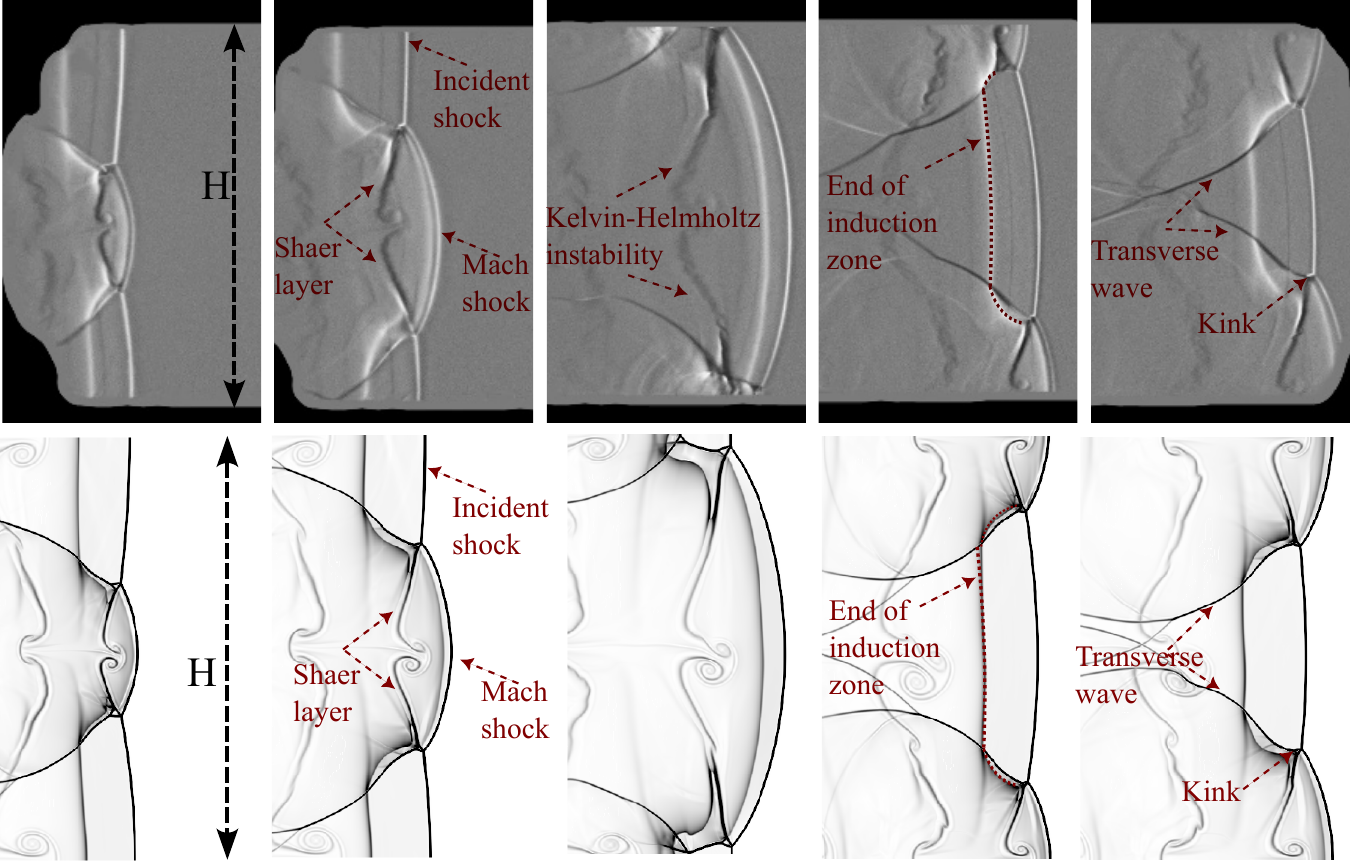}
        \caption{$\text{2H}_2/\text{O}_2/7\text{Ar}$ at $p_0 = 4.1$ \si{\kilo\pascal}}
        \label{fig:compAr}
    \end{subfigure}
    \hfill
    \begin{subfigure}[b]{\textwidth}
        \centering
	\includegraphics[width=0.9\linewidth]{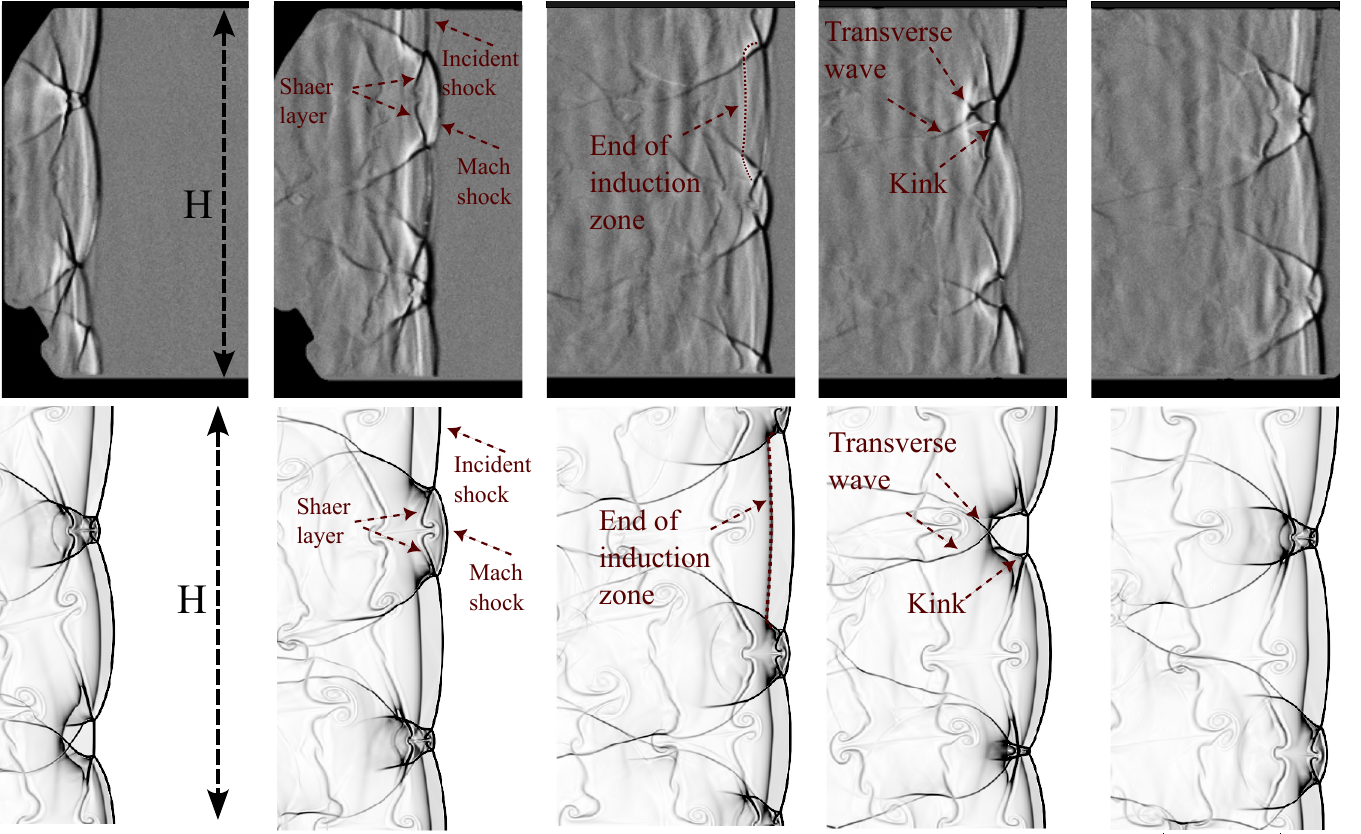}
        \caption{$\text{2H}_2/\text{O}_2/7\text{He}$ at $p_0 = 9.3$\si{\kilo\pascal}}
        \label{fig:compHe}
    \end{subfigure}
	\caption{Comparisons of the gradient of the density for experiments and simulations with $\Delta_{i,CJ}$= 2.8 \si{\milli\meter}. Second and fourth rows: the quasi-2D simulation,  first and third rows: the schlieren photos of the experiment. $H$ is the channel height of 203 mm.}
        \label{fig:compArHe}
\end{figure}
\begin{figure}[t!]
    \centering
    \begin{subfigure}[b]{0.75\textwidth}
        \centering
        \includegraphics[width=\textwidth]{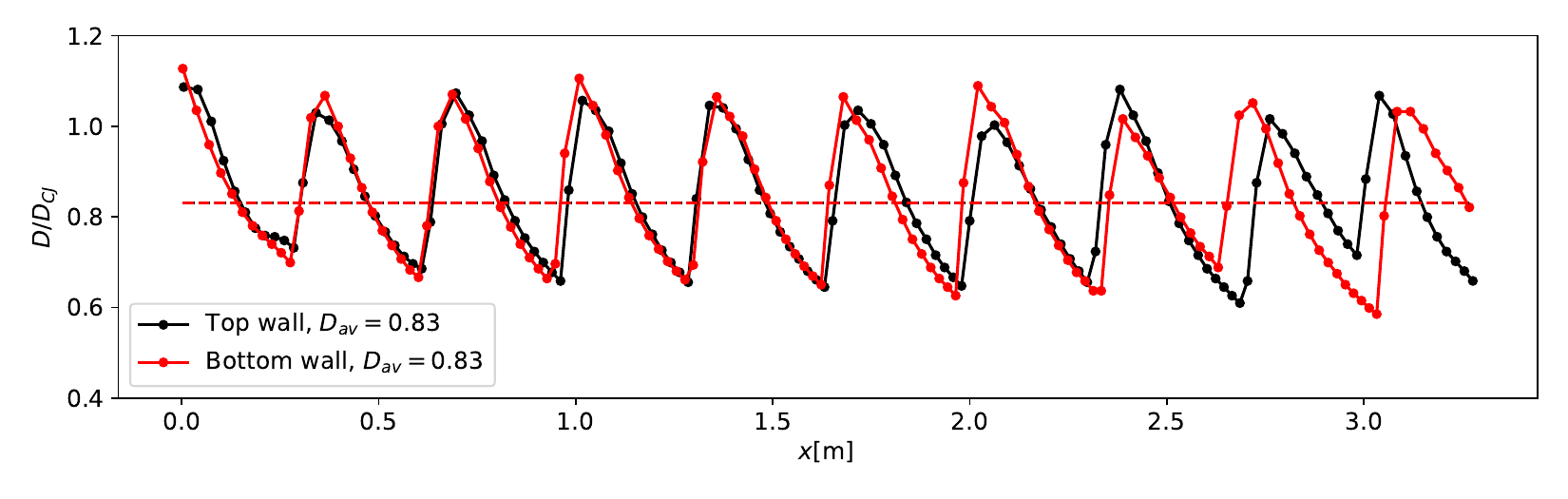}
        \caption{$\text{2H}_2/\text{O}_2/7\text{Ar}$ at $p_0 = 4.1$ \si{\kilo\pascal}}
        \label{fig:Ar-4-Shad}
    \end{subfigure}
    \hfill
    \begin{subfigure}[b]{0.75\textwidth}
        \centering
        \includegraphics[width=\textwidth]{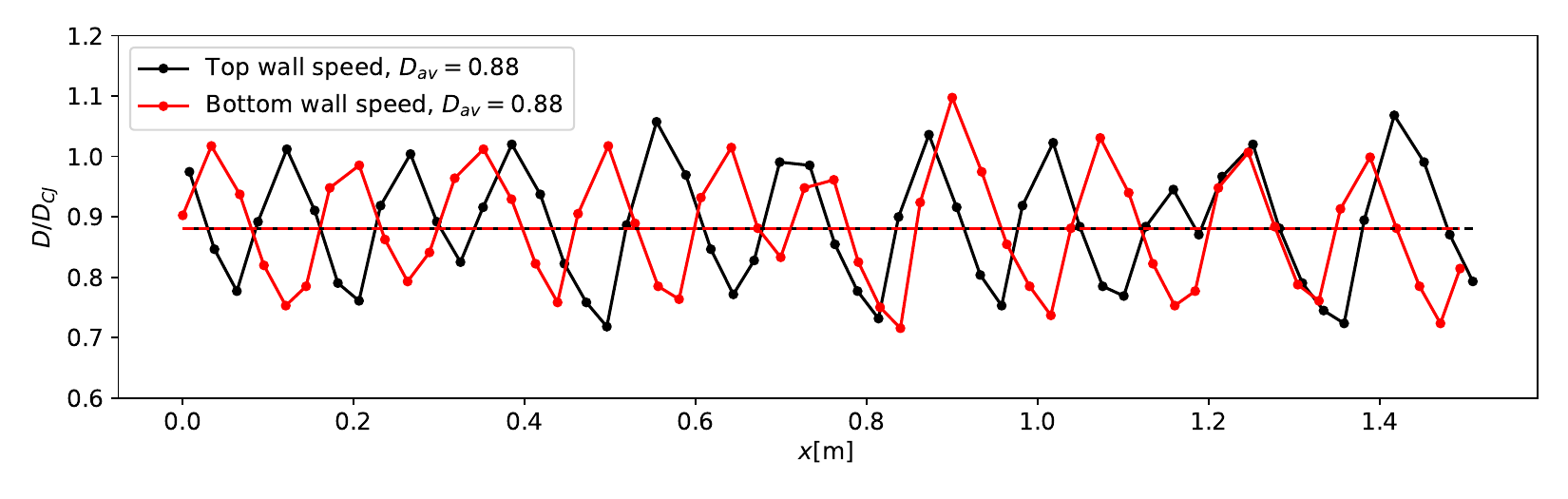}
        \caption{$\text{2H}_2/\text{O}_2/7\text{He}$ at $p_0 = 9.3$ \si{\kilo\pascal}}
        \label{fig:He-15-Shad}
    \end{subfigure}
    \caption{Shock speed evolution over the length of the ten repeated cells, extracted from the top and bottom walls of the numerical simulation. }
    \label{fig:shadSimVel}
\end{figure}
\subsection{Comparison with experiment}
Figure \ref{fig:sootAr} shows the numerically tracked maximum energy release rates for detonations in the mixtures of \(\text{2H}_2/\text{O}_2/7\text{Ar}\) at \(p_0 = 4.1\) \si{\kilo\pascal}. This corresponds to the open shutter photograph in experiments. The simulation results indicate that in the absence of losses (\(K_M = 0\)), the ideal CJ detonation exhibits five stable cells across the channel, corresponding to \(\lambda_{CJ} = 37\) \si{\milli\meter}. However, when boundary layer losses are included, using the nominal Mirels constant, \(K_M \approx 4.0\), only a single-head detonation propagates in the channel, with a higher velocity deficit than measured in the experiments. The present simulations, however, show the reduction in $K_M$ by 25\%, to a value of \(K_M = 3\) can accurately replicate the experimental results in terms of cell size and velocity deficit. Recent work by Smith et al. \cite{smith2024nature} found that using more realistic chemical model yielded good agreement with the experiment. We conclude that viscous losses capture the experiment well, within the accuracy of the known chemical kinetics and modelling assumptions.

A similar conclusion can be made from the calculations of the helium dilution case at \(p_0 = 9.3\) \si{\kilo\pascal}, as shown in Fig.\ \ref{fig:sootHe}.  The nominal \(K_M = 3.8\) results in a larger cell size and velocity deficit, while a \(K_M = 2.8\) better matches the experimental observations. 

\begin{figure}[t]
    \centering
    \begin{subfigure}[b]{0.45\textwidth}
        \centering
        \includegraphics[width=\textwidth]{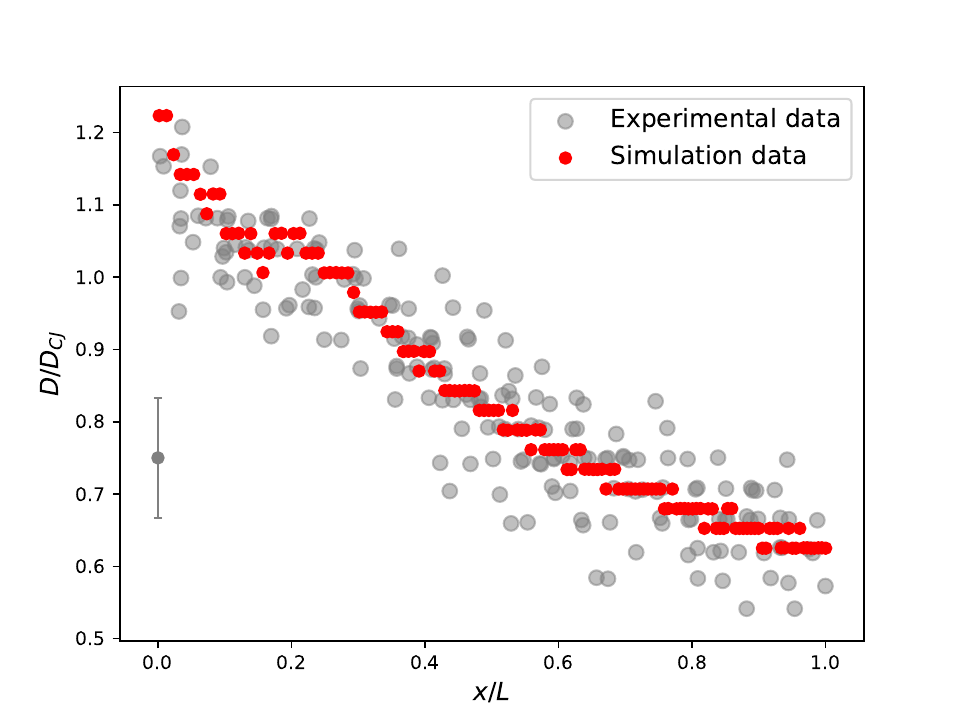}
        \caption{$\text{2H}_2/\text{O}_2/7\text{Ar}$ at $p_0 = 4.1$ \si{\kilo\pascal}}
        \label{fig:Ar-cell4-sim}
    \end{subfigure}
    \hfill
    \begin{subfigure}[b]{0.45\textwidth}
        \centering
        \includegraphics[width=\textwidth]{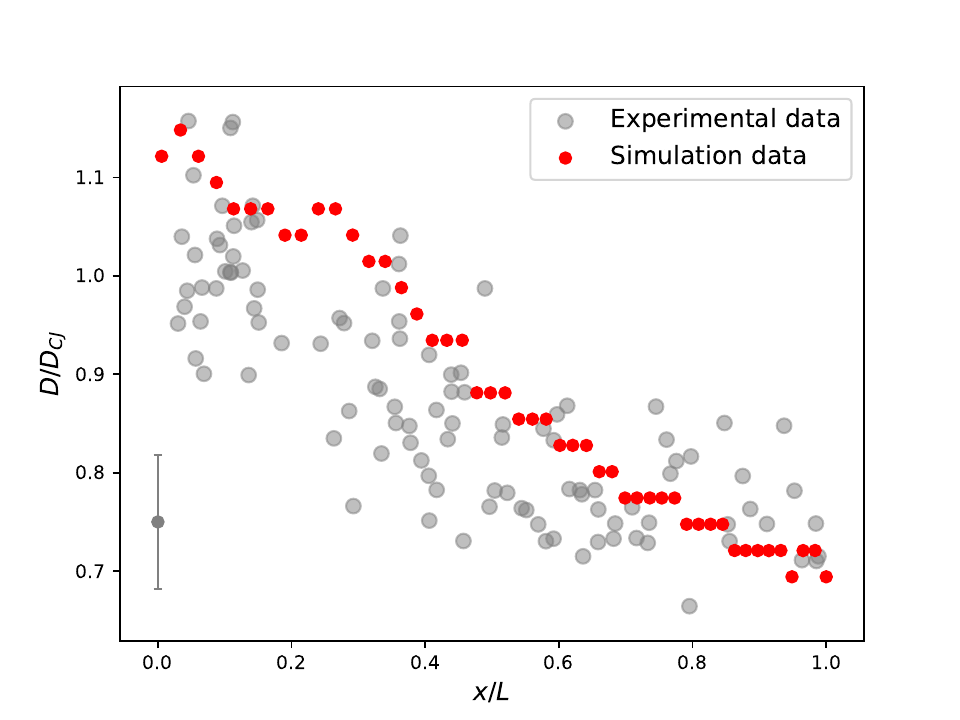}
        \caption{$\text{2H}_2/\text{O}_2/7\text{He}$ at $p_0 = 9.3$ \si{\kilo\pascal}}
        \label{fig:He-cell6-sim}
    \end{subfigure}
\caption{Comparison of shock speed evolution over a detonation cell between simulation (red circles) and experimental data (gray circles). The error bars represent the experimental uncertainty.}
    \label{fig:simonecell}
\end{figure}
It is worthwhile making a more direct comparison between the cellular structure obtained numerically and experimentally.  Figure \ref{fig:compArHe} shows evolution of the density gradients obtained in the experiments and in the simulations for the argon and helium mixtures.  The numerical simulations use the adjusted values of $K_M$.  The intricate dynamics of cellular patterns, such as unburned gas trailing the incident shock, the shear layer behind the Mach shock, and the kinks in transverse waves, are accurately reproduced when compared to experimental observations in both mixtures with the same fidelity.  This suggests the robustness of the proposed quasi-2D formulation as well as the two-step chemistry model in simulating the real detonations in experiments.  

The average detonation speed measured in ten repeated cells on the top and bottom walls of the numerical simulation is $D/D_{CJ}$=0.83 for argon diluted and 0.88 for helium diluted mixtures; these are in very good agreement with global velocity measured in the experiments (0.84 for argon diluted and 0.88 for helium diluted). Figure \ref{fig:shadSimVel} further illustrates the velocity evolution within the ten repeated cells from the top and bottom walls of the simulation, showing that steady detonation waves were achieved at the macro-scale in both mixtures, consistent with the experimental velocity profile.

We further present the temporal velocity evolution within a single cell from the 2D simulation, overlaid on the experimental data for a corresponding cell, as shown in Fig.\ \ref{fig:simonecell}. These results show good quantitative agreement between the experiment and simulation, with the maximum velocity is predicted to be 1.2 at the beginning of the cell, decreasing to 0.6 at the end for the argon-diluted mixture, and from 1.15 to 0.7 within a single detonation cell for the helium-diluted mixtures.
\subsection{The steady ZND structure and viscous scaling}
We have also compared our numerical and experimental results with the predictions made from the steady ZND model supplemented by the lateral flow losses described above. This model permitted us to determine the loss parameter that can permit scaling the results with the different diluents.  

The steady wave structure is evaluated in the framework of one-dimensional ZND detonations with flow divergence to the boundary layers \cite{zangene2021role, radulescu2018dynamics}.  The Euler governing equations for the steady, inviscid, reactive quasi-1D flow in the shock-attached reference frame are:
\begin{equation}
\frac{dp}{dt'} = -\rho u'^2 \frac{\dot{\sigma} - \dot{\sigma}_A}{1 - M^2}
\label{eq:dpdt'}
\end{equation}

\begin{equation}
\frac{du'}{dt'} = u' \frac{\dot{\sigma} - \dot{\sigma}_A}{1 - M^2}
\label{eq:dudt'}
\end{equation}

\begin{equation}
\frac{d\rho}{dt'} = -\rho \frac{\dot{\sigma} - M^2 \dot{\sigma}_A}{1 - M^2}
\label{eq:drhodt'}
\end{equation}
\begin{equation}
\frac{dx'}{dt'} = u'
\label{eq:dxdt'}
\end{equation}
where \( p \), \( \rho \), and \( u'\) are the mixture pressure, density, and the post-shock flow velocity in the shock-attached frame of reference, respectively, and \( M \) is the Mach number of the flow. For a mixture of ideal gases, the thermicity reduces to the equation:
\begin{equation}
\dot{\sigma} = \sum_{i=1}^{N_s} \left(\frac{W}{W_i} - \frac{h_i}{c_p T}\right) \frac{dY_i}{dt'}
\label{eq:thermicity}
\end{equation}
in this equation, \( W_i \), \( Y_i \), and \( h_i \) are the molecular weight, mass fraction, and enthalpy of the \( i^{th} \) species, respectively; \( W \) is the mean molecular weight, \( c_{p} \) is the specific heat, and \( T \) is the temperature of the mixture. The kinetics for the evolution of the mass fractions, \( Y_i \), of each species is given by:
\begin{equation}
\frac{dY_i}{dt'} = \frac{W_i \dot{\omega}_i}{\rho}
\label{eq:dydt'}
\end{equation}
where \( \dot{\omega}_i \) is the molar production rate of species \( i \). The lateral strain rate of the flow in the channel is calculated by:
\begin{equation}
\dot{\sigma}_A = u' \frac{1}{A} \frac{dA}{dx'}
\label{eq:dsigmaadt'}
\end{equation}
the term $\frac{1}{A} \frac{dA}{dx'}$ was previously defined in Eq.\ \eqref{eq:source}
\begin{figure}[t!]
    \centering
    \begin{subfigure}[b]{0.45\textwidth}
        \centering
        \includegraphics[width=\textwidth]{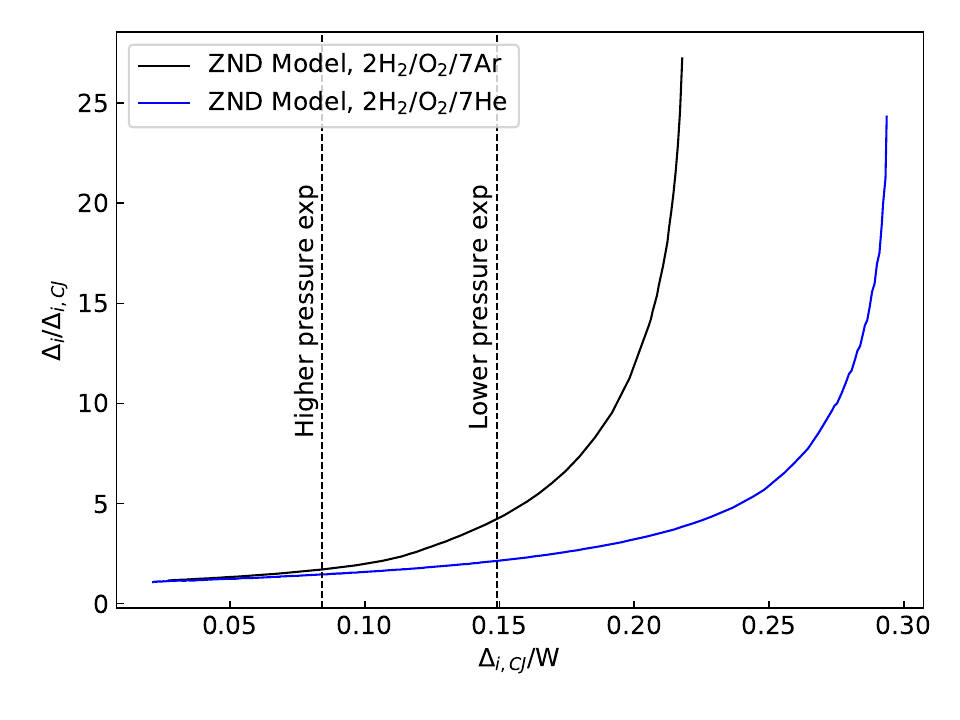}
        \caption{}
        \label{fig:ZND-1}
    \end{subfigure}
    \hfill
    \begin{subfigure}[b]{0.45\textwidth}
        \centering
        \includegraphics[width=\textwidth]{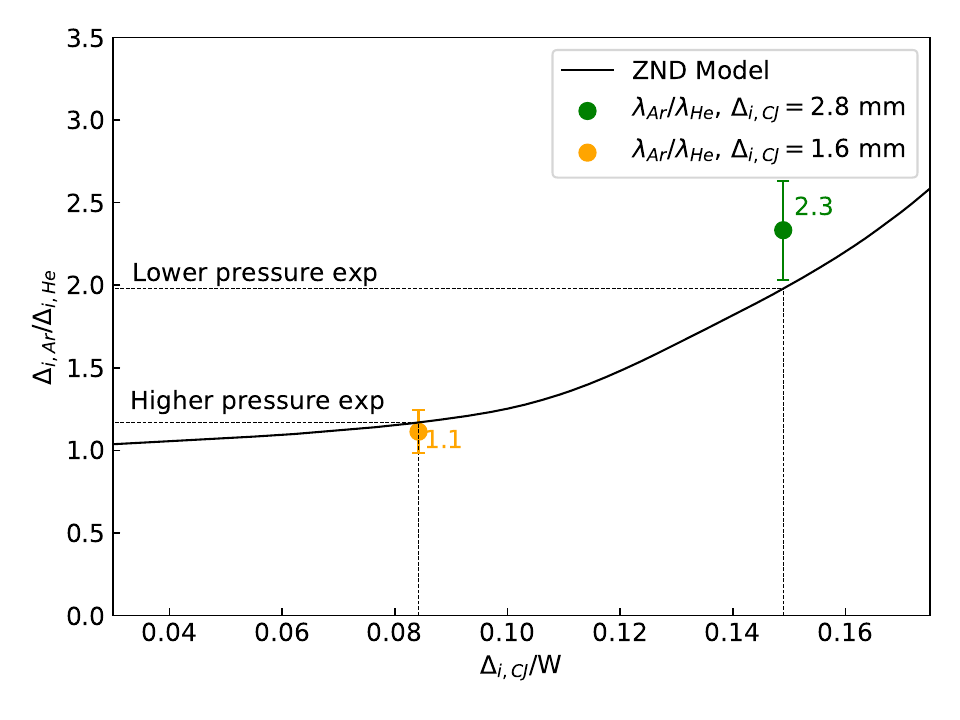}
        \caption{}
        \label{fig:ZND-2}
    \end{subfigure}
    \caption{a) The evolution of the non-dimensional induction zone length from the ZND model with losses versus the inverse of the channel width, $W$. b) The ratio of the induction zone length for argon-diluted and helium-diluted mixtures from the ZND model with losses versus the inverse of the channel width. $\lambda$ represents the cell width ratio obtained from experiments (Table \ref{tab:Exp}) and $\Delta_{i,CJ}$ is the channel width, measuring 19 \si{\milli\meter}.}
    \label{fig:ZNDLoss}
\end{figure}
A custom Python code was developed for the real gas calculations, utilizing Cantera's framework to compute the  thermo-chemical data \cite{cantera} and the Shock and Detonation Toolbox \cite{SDToolbox} to evaluate the shock jump conditions. The system of ODEs was then solved using SciPy’s variable-coefficient ODE solver (VODE). For each mixture and initial state, the detonation speed and reaction zone structure were obtained by a shooting method for the speed eigenvalue. The correct speed was such that the generalized CJ condition was satisfied inside the reaction zone, where the flow becomes sonic as the rate of energy release ($\dot{\sigma}$) balances the rate of the loss ($\dot{\sigma}_A$). We measured the induction zone length using the ZND model with boundary layer losses ($\Delta_{i}$) and without them ($\Delta_{i,CJ}$), as shown in Fig.\ \ref{fig:ZNDLoss}. In Fig.\ \ref{fig:ZND-1}, the black line represents the argon-diluted mixture, while the blue line corresponds to the helium-diluted mixture. The divergence between the two lines increases as the channel thickness decreases, consistent with the observed differences in cell sizes between the two mixtures, particularly in lower pressure experiments. For a clearer comparison, Fig.\ \ref{fig:ZND-2} presents the ratio of the induction zone length of the argon-diluted mixture to that of the helium-diluted mixture, as calculated by the ZND model with losses, alongside the experimentally measured cell widths. The orange circle is the cell width ratio in both argon and helium dilution in the higher-pressure experiment and the green circle is related to the same ratio in the lower-pressure experiments. The error bars indicate the standard deviation calculated from repeated experiments conducted in the same mixture under the same initial pressure conditions. This graph shows that the ZND model with losses can predict the experimental result with good accuracy. This result explains why the cell width in the lower-pressure experiments is not precisely the same, and it is because of the higher losses near the limit.
\begin{figure}[t!]
    \centering
    \begin{subfigure}[b]{0.45\textwidth}
        \centering
        \includegraphics[width=\textwidth]{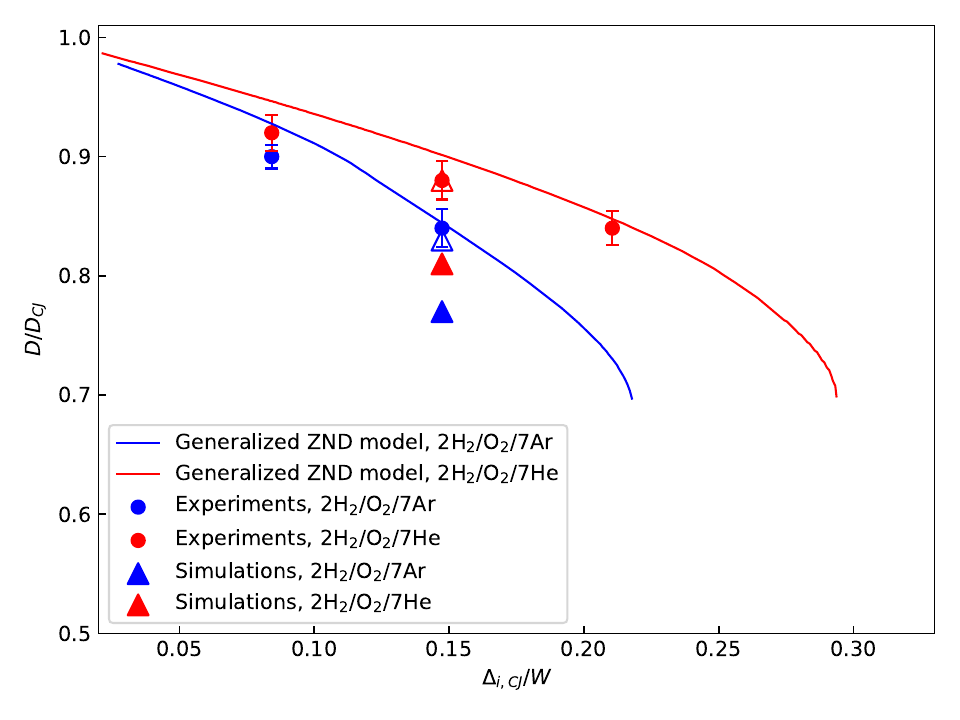}
        \caption{}
        \label{fig:velocity}
    \end{subfigure}
\begin{subfigure}[b]{0.45\textwidth}
        \centering
        \includegraphics[width=\textwidth]{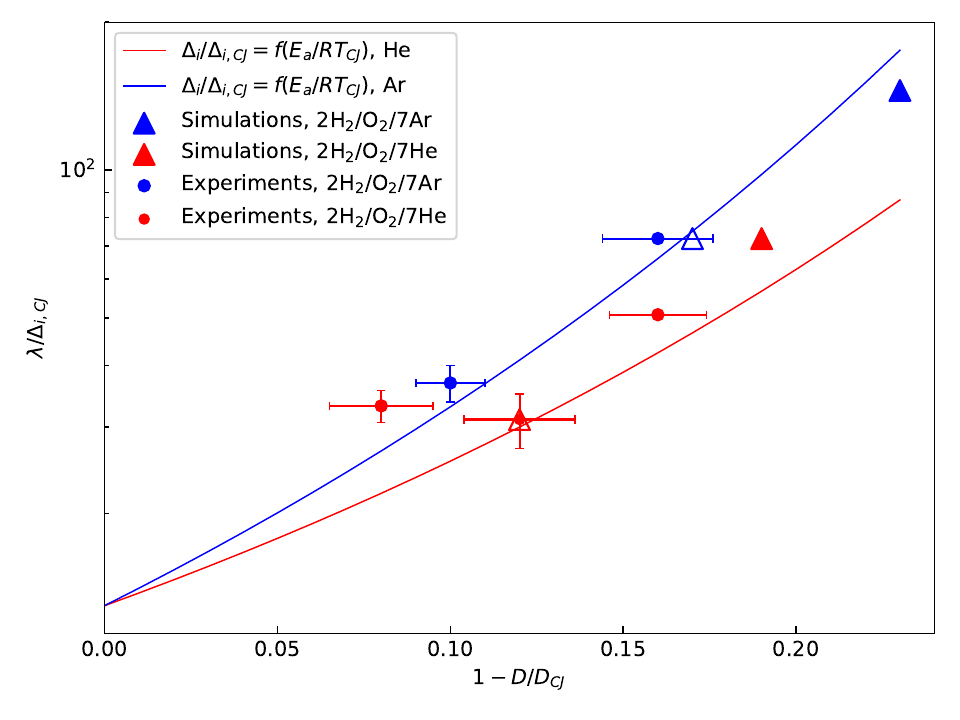}
        \caption{}
        \label{fig:deltai}
    \end{subfigure}
\hfill
	\caption{a) The mean propagation velocity in argon and helium mixtures measured experimentally (symbols) and calculated from the ZND model with losses (lines); the abscissa is the inverse channel thickness normalized by the induction zone length at CJ condition. b) Comparisons of the measured velocity deficit and cell sizes from quasi-2D simulation with experiments. The solid line is the sensitivity of induction kinetics with temperature.  Closed triangles represent the simulation results using the nominal \(K_M\) values (4 for the argon mixture and 3.8 for the helium mixture), while open triangles indicate the results with adjusted values (3 for the argon mixture and 2.8 for the helium mixture).}
	\label{fig:deficit}
\end{figure}

Figure \ref{fig:velocity} presents the numerically calculated velocity deficits for both mixtures as a function of the non-dimensionalized CJ induction zone length, alongside velocity deficits measured from experiments and 2D simulations. Overall, considering the experimental, simulation, and ZND model data, the velocity deficits in the lower-pressure experiments (\(\Delta_{i,CJ}/W = 0.15\)) are consistently higher compared to those in the higher-pressure experiments (\(\Delta_{i,CJ}/W = 0.084\)). As shown in the figure, the adjusted values closely match the experimental data, while the nominal values alter the velocity deficit prediction by 8\%. Although the ZND model does not predict the experimental results as accurately as the 2D simulations, its predictions still fall within the experimental uncertainty. The discrepancy between the 2D simulations and the ZND model may be attributed to the influence of the cellular structure.

Figure \ref{fig:deltai} illustrates the dimensionless relationships, represented by the solid line, that describe the velocity deficit in relation to the detonation cell size ($\lambda$) and the CJ induction zone length ($\Delta_{i, CJ}$). From the 2D simulation results, we found that $\frac{\lambda_{CJ}}{\Delta_{i,CJ}}=13$ in the lower-pressure experiments where $\Delta_{i,CJ}=2.8$ \si{\milli\meter} and $\lambda_{CJ}=37$ \si{\milli\meter}. Using Eq. 12 from \cite{Xiao:2021}, $\frac{\Delta_{i}}{\Delta_{i,CJ}} \approx \frac{\lambda}{\lambda_{CJ}} \approx \exp\left(\frac{2E_a}{RT_{CJ}}(1 - \frac{D}{D_{CJ}})\right)$, we compute $\frac{\lambda}{\lambda_{CJ}} \times \frac{\lambda_{CJ}}{\Delta_{i,CJ}} = \frac{\lambda}{\Delta_{i,CJ}}$, which are displayed as solid lines in the figure. The adjusted 2D simulation accurately predicts both the cell size and velocity deficit observed in the both mixtures. However, using the nominal values of \(K_{M}\) results in an overestimation of the cell sizes by a factor of two.

To summarize, the overall comparison between experiment and simulations suggest that velocity deficits, cell size and overall reaction structure can be well captured by a quasi-2D CFD model with incorporation of viscous losses.  Perfect agreement can be obtained by modifying one of the model closed parameters by 25 \%.  This margin of error has been claimed to be reduced by Smith et al. \cite{smith2024nature}, by using a higher fidelity thermo-chemical model.  This success is quite remarkable, and suggests the viscous losses dominate the observed effects.

\subsection{Theoretical viscous scaling law}

\begin{figure}[t!]
    \centering
        \includegraphics[scale=0.6]{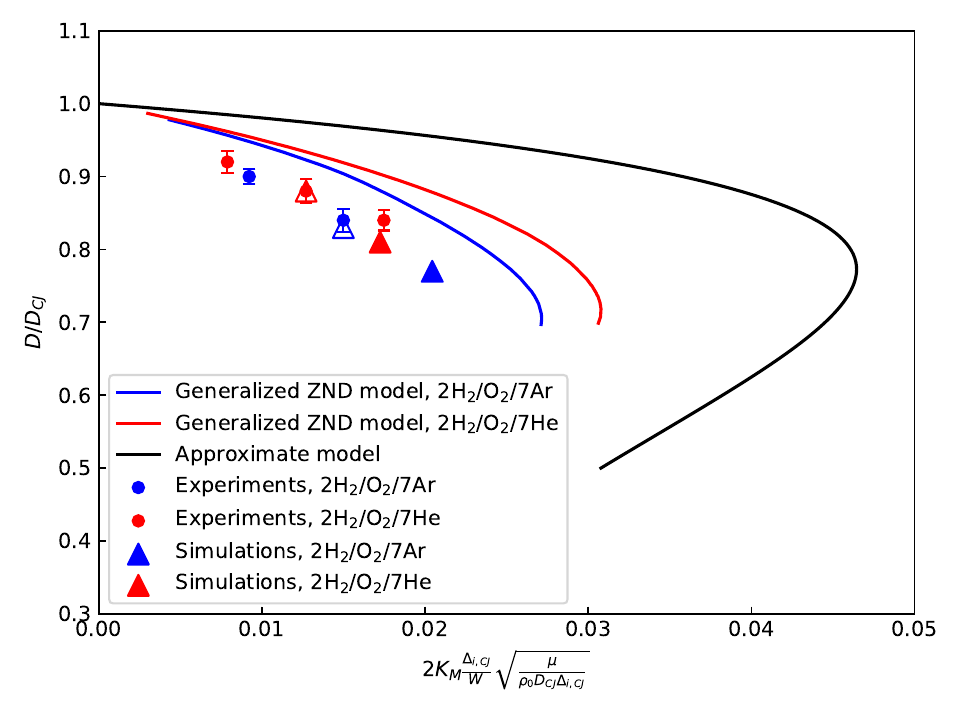}
	\caption{The evolution of non-dimensional detonation speeds as a function of the loss term, compared with experimental data, 2D simulations, and the generalized ZND model. Closed triangles represent the simulation results using the nominal \(K_M\) values (4 for the argon mixture and 3.8 for the helium mixture), while open triangles indicate the results with adjusted values (3 for the argon mixture and 2.8 for the helium mixture).}
	\label{fig:aprox}
\end{figure}

A detailed analysis of the response of detonations to lateral flow divergence due to laminar boundary layers is presented in the appendix.  We conducted a perturbation analysis in the limit of a high activation energy, assuming a square wave detonation, following the main ideas of He and Clavin \cite{he1994direct}. The result of this analysis shows that the velocity deficit parameter 
\begin{equation}
\phi=\frac{D-D_{CJ}}{\epsilon D_{CJ}}
\end{equation}
depends on the loss parameter
\begin{equation}
\zeta = \frac{4\gamma^2}{\gamma+1} \frac{1}{\epsilon}\frac{\Delta_{i,CJ}}{W} K_M \sqrt{\frac{\mu}{\rho_0 D_{CJ} \Delta_{i,CJ}}}
\end{equation}
by 
\begin{equation}
\phi e^{-\phi}=\zeta
\label{eq:loss}
\end{equation}
where \(\epsilon\) denotes the inverse of the non-dimensional activation energy. This has the characteristic inverted C shape with a maximum loss parameter permitting self-sustained propagation given by $\zeta^*=e^{-1}$.  The results of Eq.\ \eqref{eq:loss} using \(\gamma = 1.5\) and \(\frac{1}{\epsilon} = 4.4\) for the mixtures of this study is the C-shaped dashed line in Fig.\ \ref{fig:aprox}, which illustrates the relationship between the velocity deficit and the loss term. 
The results of the analysis and the identification of the loss parameter now permits us to collapse the experimental and numerical data for the two mixtures.  The successful scaling relies on the square root of the inverse Reynolds number $\sqrt{\frac{\mu}{\rho_0 D_{CJ} \Delta_{i,CJ}}}$ in the loss parameter $\zeta$.  Fig.\ \ref{fig:aprox} shows that the results of the experiments collapse on the same curve under viscous loss scaling.  This is perhaps the most convincing argument reconciling the results in argon and helium under viscous scaling.  This good scaling suggests that possible modifications due to non-equilibrium effects are negligible, as they would otherwise break this scaling.

While experiments in helium and argon collapse within the accuracy of the speed measurement, the helium data appears to indicate slightly higher speeds, hence more sensitive mixtures.  This again is not compatible with the vibration relaxation explanation, as the reactants in the presence of helium relax faster and the kinetics would not rely on the speed up due the excess translational temperature in the induction zone. While empirically, we show that the vibrational effect is negligible, this is further supported by the time scale analysis of section \ref{sec:Vib}. 

One can note, however, that the ideal model is not in quantitative agreement with the ZND model obtained numerically using the full chemistry.   In the approximate model, due to the assumption of a square wave detonation, the turning point is larger by a factor of two compared to the generalized ZND model. This behaviour is consistent with the analysis of Short and Bdzil for a closely analogous problem, who have found that the rate of energy release in the reaction zone progressively shifts the turning point at smaller losses and larger velocity deficits as the rate of energy release is lowered \cite{short2003propagation}. A more detailed model incorporating a finite reaction zone will be communicated in a sequel.  

In closing, while we did not find that vibrational relaxation to play an important role in the H$_2$/O$_2$ system, it does not imply its effect as being negligible in other fuels or other operating conditions.  For example, the empirical evidence of substantial differences between argon and helium dilution in the experiments of Haloua et al.\ in propane calls for a closer scrutiny in hydrocarbons \cite{haloua2000characteristics}. 

\section{Conclusion}\addvspace{10pt}
\label{sec:Con}
The reaction zone cellular structure of $\text{2H}_2/\text{O}_2$ detonation was found to be affected by the type of mono-atomic diluent when propagating in thin channels.  At the same nominal induction length and nearly identical chemical kinetics, the experiments of detonation propagation with 70\% argon dilution were found to yield cell sizes larger by a factor of 2 than those with 70\% helium dilution, while the velocity deficits were also larger for the argon mixture in lower pressure experiments.  These large differences permitted us to establish the importance of viscous losses and non-equilibrium effects. Non-equilibrium effects alone cannot reconcile these effects and would predict the opposite trend.  We found that viscous losses can entirely account for the observed effects.  Vibrational non-equilibrium effects suggested in the past to play a large role on the cellular dynamics were found to be negligible in the system studied.

\section*{Acknowledgments} 
M. I. Radulescu acknowledges the financial support provided by the Natural Sciences and Engineering Research Council of Canada (NSERC) through the Discovery Grant ”Predictability of detonation wave dynamics in gases: experiment and model development”.  This work was also supported by AFOSR grant FA9550-23-1-0214, with Dr.\ Chiping Li as program monitor. The authors wish to thank Dr. Q. Xiao, Dr. Z. Hong, for their valuable discussions, and Dr. A. Sow for his support with the computational code.

\section*{Supplementary material}
Supplementary material associated with this article can be found in the online version.

\bibliography{RefWFD2023}
\clearpage
\renewcommand{\theequation}{\arabic{equation}} 
\setcounter{equation}{0} 

\input{Appendix-v4}
\end{document}

%% file: Appendix-v4.tex
\section*{\Large Appendix: Detonations with Boundary Layer Losses: Square Wave Detonation}
\label{section:appendix}
We analysed the structure and propagation of a detonation wave in the presence of boundary layer losses, following the analysis of He and Clavin for cylindrical quasi-steady detonations \cite{he1994direct}. We perturb the ZND solution by a small perturbation in shock speed, which in turns entails small perturbations in the post shock state in the induction zone.  The analysis reveals that the perturbations are of the order of the inverse activation energy. 

In the shock-attached reference frame, the conservation laws for quasi-1D flow are:
\begin{equation} \label{eq:1}
\frac{d}{dx} (\rho u) = - \rho u \frac{1}{A} \frac{dA}{dx}
\end{equation}
\begin{equation} \label{eq:2}
\frac{d}{dx} (\rho u^2 + p) = - \rho u^2 \frac{1}{A} \frac{dA}{dx}
\end{equation}
\begin{equation} \label{eq:3}
\frac{d}{dx} \left( h + \lambda Q + \frac{1}{2} \rho u^2 \right) = 0
\end{equation}
here, \(p\), \(\rho\), \(u\), \(h\), \(A\) and \(Q\) denote pressure, density, flow velocity in the x-direction, enthalpy, cross-section area and the chemical energy release. In the induction zone:
\begin{equation}\label{eq:4}
u\frac{d\xi}{dx} = K_i \exp \left( - \frac{E_a}{RT} \right)
\end{equation}
with  \(\xi = 0\) at the shock and \(\xi = 1\) marking the end of induction zone. In the reaction zone:
\begin{equation}\label{eq:5}
 u\frac{d \lambda}{dx} = - \frac{\omega}{\rho} = - K_r \lambda
\end{equation}
here we take \(\lambda = 1\) in the reactants and \(\lambda = 0\) in the products. We will assume \(K_r\) to be very large, such that the structure is a square wave \cite{short2003propagation}. In the limit of \(K_r \to \infty\), \( \lambda \to 0\) in the burned gases.

\begin{figure}[t]
\centering
\includegraphics[scale=0.35]{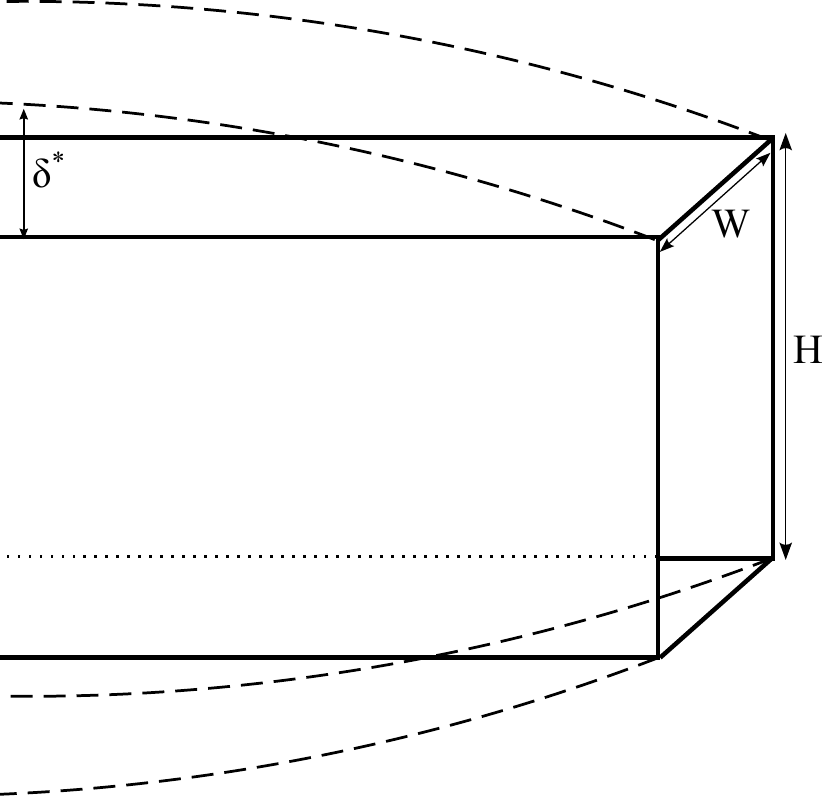}
\caption{Boundary layer displacement thickness in a rectangular channel.}
\label{fig:bl}
\end{figure}

\begin{figure}[t]
\centering
\includegraphics[scale=0.82]{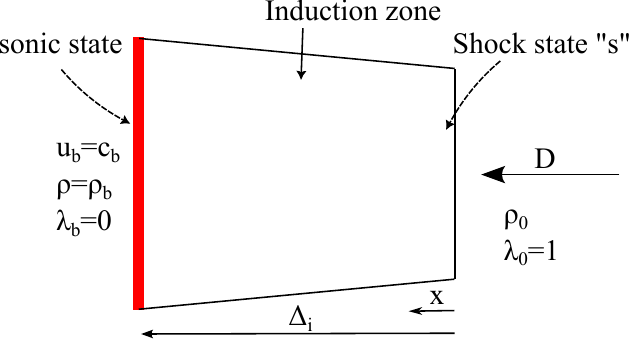}
\caption{Sketch illustrating the square wave detonation model.}
\label{fig:sq}
\end{figure}

The right hand side of the governing equations contain the nozzling term involving the lateral strain rate.  These terms are higher order, and hence only their leading forms must be retained.  We will assume a rectangular channel with effective area of $A = (W + 2 \delta^*) (H + 2 \delta^*)$ as shown in Fig. \ref{fig:bl}, where $\delta^*$ is the negative displacement thickness due to laminar boundary layers obtained by Mirels, $\delta^* = K_M \sqrt{\frac{\mu x}{\rho u}}$. 
The nozzling term can thus be written as:
\begin{equation}\label{eq:6}
\alpha = \frac{1}{A} \frac{dA}{dx} \simeq K_M\left( \frac{1}{W} + \frac{1}{H} \right) \sqrt{\frac{\mu}{\rho u x}}
\end{equation}
We can then dispense with integrating the equations in the very thin reactive zones, and simply use the boundary condition at the sonic surface (\( u_b = c_b \), \( \lambda_b = 0 \)), shown in Fig. \ref{fig:sq}. 
\begin{equation}
\rho_b u_b = \rho_0 D - \int_{0}^{\Delta_i} \rho u 
\left(\frac{1}{W} + \frac{1}{H}\right) K_M 
\sqrt{\frac{\mu}{\rho u}} x^{-\frac{1}{2}} \, dx
\label{eq:17}
\end{equation}
\begin{equation}
\rho_b u_b^2 + p_b = \rho_0 D^2 + p_0 - 
\int_{0}^{\Delta_i} \rho u^2 \left(\frac{1}{W} + \frac{1}{H}\right) K_M  
\sqrt{\frac{\mu}{\rho u}} x^{-\frac{1}{2}} \, dx
\label{eq:18}
\end{equation}
\begin{equation}
\frac{\gamma}{\gamma - 1} \frac{p_b}{\rho_b} + \frac{1}{2} u_b^2 = 
\frac{\gamma}{\gamma - 1} \frac{p_0}{\rho_0} + Q + \frac{1}{2} D^2
\label{eq:19}
\end{equation}
For simplicity, we will deal with strong detonations, such that terms including \(p_0\) in momentum equation and \(\frac{\gamma}{\gamma -1}\frac{p_0}{\rho_0}\) in energy equation will be negligible. To integrate the integrals in Eq.\ \eqref{eq:17} and Eq.\ \eqref{eq:18}, we will assume that the variables in the induction zone can be written as:
\[
\rho = \rho_s + \rho', \quad u = u_s + u', \quad \dots
\]
with $\rho'$, $u'$ being small perturbations. Re-write Eq.\ \eqref{eq:17} by only retaining leading order terms in the right-hand side and integrate:
\begin{equation}
\rho_b u_b = \rho_0 D \left(1 - 2 \left(\frac{1}{W} + \frac{1}{H}\right) K_M  
\sqrt{\frac{\mu \Delta_i}{\rho_0 D}} \right)
\label{eq:20}
\end{equation}
where we used $\rho_s u_s = \rho_0 D$ from mass conservation across the shock. The momentum equation, Eq.\ \eqref{eq:18}, can be written in a similar way for the leading order perturbations
\begin{equation}
\rho_b u_b^2 + p_b = \rho_0 D^2 \left(1 -2 \frac{\gamma - 1}{\gamma + 1}
\left(\frac{1}{W} + \frac{1}{H}\right) K_M  \sqrt{\frac{\mu \Delta_i}{\rho_0 D}} \right)
\label{eq:21}
\end{equation}
in obtaining Eq.\ \eqref{eq:21}, we have used the strong shock relation of $\frac{u_s}{D} = \frac{\rho_0}{\rho_s} = \frac{\gamma - 1}{\gamma + 1}$. The momentum, Eq.\ \eqref{eq:21}, and energy, Eq.\ \eqref{eq:19}, equations can be re-written by eliminating $p_b$ using the sonic condition ($c_b^2 = \frac{\gamma p_b}{\rho_b} = u_b^2$). Re-write the mass, momentum, and energy equations:
\begin{equation}
\rho_b u_b = \rho_0 D \left(1 - \beta \left(\frac{\Delta_i}{\Delta_{i,CJ}}\right)^{\frac{1}{2}} 
\left(\frac{D}{D_{CJ}}\right)^{-\frac{1}{2}}\right)
\label{eq:25}
\end{equation}
\begin{equation}
\frac{\gamma + 1}{\gamma} \rho_b u_b^2 = \rho_0 D^2 \left(1 - \frac{\gamma - 1}{\gamma + 1} 
\beta \left(\frac{\Delta_i}{\Delta_{i,CJ}}\right)^{\frac{1}{2}} \left(\frac{D}{D_{CJ}}\right)^{-\frac{1}{2}}\right)
\label{eq:26}
\end{equation}
\begin{equation}
\frac{\gamma + 1}{2(\gamma - 1)} u_b^2 = Q + \frac{1}{2} D^2
\label{eq:27}
\end{equation}
with
\[
\beta = 2\left(\frac{1}{W} + \frac{1}{H}\right) \Delta_{i,CJ} K_M  \sqrt{ \frac{\mu}{\rho_0 D_{CJ}\Delta_{i,CJ}}}
\]
note that $\beta$ is the constant parameter denoting the magnitude of the losses. We can now proceed to eliminate $\rho_b$ and $u_b$ from the system of equations Eq.\ \eqref{eq:25}-\eqref{eq:27}, for $\beta \ll 1$ the following expression can be obtained:
\begin{equation} \label{eq:33}
\frac{\gamma^2}{\gamma^2 - 1} \left( \frac{D}{D_{CJ}} \right)^2 \left[ 1 + \frac{4}{\gamma + 1} \beta \left( \frac{\Delta_i}{\Delta_{i,CJ}} \right)^{\frac{1}{2}}  \left( \frac{D}{D_{CJ}} \right)^{-\frac{1}{2}} \right] = 2 \frac{Q}{D_{CJ}^2} + \left( \frac{D}{D_{CJ}} \right)^2
\end{equation}
To close the system, we need an expression for $\frac{\Delta_i}{\Delta_{i,CJ}}$ in terms of $\frac{D}{D_{CJ}}$.
The induction zone length is obtained by integrating Eq.\ \eqref{eq:4}, to the leading order:
\[
    \Delta_i = \frac{u_s}{k_i} \exp \left(\frac{E_a}{R(T_s)}\right)
\]
Similarly, for the nominal case without losses:
\[
    \Delta_{i,CJ} = \frac{u_{s,CJ}}{k_i} \exp \left(\frac{E_a}{R(T_s,CJ)}\right)
\]
For strong shocks, $u_s \sim D$ and $T_s \sim D^2$, therefore:
\begin{equation} \label{eq:34}
   \begin{aligned}
      \frac{\Delta_i}{\Delta_{i,CJ}}  &= \frac{D}{D_{CJ}} \exp \left( \frac{E_a}{R(T_s,CJ)} \left( \frac{1}{(D/D_{CJ})^2} - 1 \right) \right)
   \end{aligned}
\end{equation}
replacing Eq.\ \eqref{eq:34} in the equation, Eq.\ \eqref{eq:33} and assuming small perturbations of speed from the CJ state, a relation between $\left( \frac{D}{D_{CJ}} \right)^2$ and the losses is obtained:
\begin{figure}[t]
\centering
\includegraphics[scale=0.6]{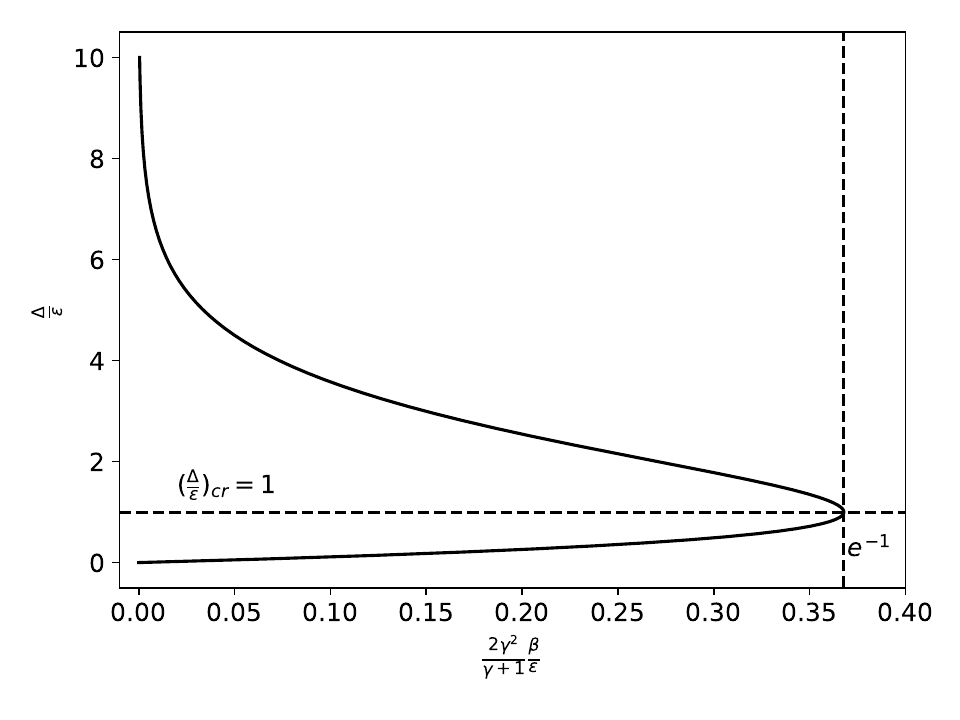}
\caption{The dual solution C-curves for boundary layer induced detonation quenching.}
\label{fig:turning}
\end{figure}
\begin{equation} \label{eq:37}
    \frac{\Delta}{\varepsilon} \exp\left(-\frac{\Delta}{\varepsilon}\right) = \frac{2\gamma^2}{\gamma+1} \frac{\beta}{\varepsilon}, 
    \quad \varepsilon = \left(\frac{E_a}{RT_{s,CJ}}\right)^{-1}
\end{equation}
Figure \ref{fig:turning} shows the dual-solution behaviour given by Eq.\ \eqref{eq:37}. The turning point will be at the zero slope of $ \frac{2 \gamma^2}{\gamma +1 } \frac{\beta}{\varepsilon}$ in terms of the argument of the exponential. Differentiate the LHS of Eq.\ \eqref{eq:37}, the maximum loss term is:
\begin{equation} \label{eq:39}
    \beta^* = \frac{\gamma + 1}{2 \gamma^2} \frac{\varepsilon}{e}
\end{equation}
where $ e^{-1} = \frac{\gamma^2}{\gamma + 1} \frac{\beta^*}{\varepsilon}$. As expected, $\beta^*$ is a small quantity of order $\varepsilon$.

In summary, Eq.\ \eqref{eq:37} provides the dependence of detonation speed on the loss parameter, with the maximum loss permitting propagation given by Eq.\ \eqref{eq:39}.